\documentclass[fleqn,usenatbib]{mnras}

\usepackage[T1]{fontenc}

\DeclareRobustCommand{\VAN}[3]{#2}
\let\VANthebibliography\thebibliography
\def\thebibliography{\DeclareRobustCommand{\VAN}[3]{##3}\VANthebibliography}

\usepackage{graphicx}	
\usepackage{amsmath}	
\usepackage{subfigure} 

\title[Detailed analysis of variable stars in NGC 7006]{Detailed analysis of the variable star population in the globular cluster NGC 7006.}

\author[A. Arellano Ferro et al.]{
A. Arellano Ferro,$^{1}$\thanks{E-mail: armando@astro.unam.mx}
F. C. Rojas Galindo,$^{1}$\thanks{E-mail: fcrojas@astro.unam.mx}
I. H. Bustos Fierro,$^{2}$
S. Muneer$^{3}$ \and
M.A. Yepez,$^{1}$ 
Sunetra Giridhar$^{3}$
\\
$^{1}$Instituto de Astronomía, Universidad Nacional Autónoma de México, Ciudad Universitaria, C.P. 04510, México\\
$^{2}$Observatorio Astronómico, Universidad Nacional de Córdoba, Córdoba C.P. 5000, Argentina \\
$^{3}$Indian Institute of Astrophysics, Sarjapur Main Road, 2nd Block, Koramangala 560034, Bangalore, India
}

\date{Accepted XXX. Received YYY; in original form ZZZ}

\pubyear{2021}

\begin{document}
\label{firstpage}
\pagerange{\pageref{firstpage}--\pageref{lastpage}}
\maketitle

\begin{abstract}
A five year CCD photometric \emph{VI} time-series of NGC 7006 is employed to perform a detailed analysis of the known population of variable stars. In the process we have corrected inconsistent classifications, sky coordinates and found ten new cluster member variables. An independent reddening estimate with a value $E(B-V)=0.08\pm0.05$ is made. Using Fourier decompositions of RR Lyrae light curves and well established calibrations, the cluster mean metallicity and distance [Fe/H]$_{\rm ZW}= -1.53\pm0.15$ and $41.2\pm1.4$ kpc are estimated based on an extended sample of cluster member RRab stars. Using the $Gaia$-DR3 data we performed an extensive membership analysis which leads to a clean Colour-Magnitude diagram, and hence to the identification of variables that are likely field stars, and to considerations on the variables distribution in the Horizontal Branch (HB). A double mode RR Lyrae and three CW star are discussed. The origin of CW stars from precursors in the blue tail of the HB with very thin ($\sim 0.06 \pm 0.01 M_{\odot}$) envelopes is argued. Our models indicate that the Main Sequence predecessor of RR Lyrae stars had a mass of 0.82-0.85 $M_{\odot}$ and lost about 25-35\% of its mass during the red giant branch events before settling in the HB some 12-13.5 Gyrs later. 
\end{abstract}




\section{Introduction}

The study of the variable star populations in globular clusters from CCD time-series imaging has proven to be successful in the determination of physical parameters such as the metallicity and distance, and in the detailed description of the Color Magnitude diagram structure. The approach has also enabled us to complete the variable star population census of the cluster under study, to improve upon the pulsation periods and to correct variable star type classifications, to confirm the variable vs non-variable nature of many misclassified stars and often to correct sky coordinates and to accurately identify all variables star in the field of the cluster. Furthermore, the recent publication of the results of the $Gaia$ mission, the DR3 release, and the development of powerful statistical numerical approaches, offer an unprecedented opportunity to separate the cluster members from the field stars, which leads to more accurate Colour-Magnitude diagrams (CMD), and particularly the distribution of member variables stars and their confrontation with their evolutionary properties. For all the above reasons there is a renewed interest in depth study of dense time series of images of globular clusters.

The present work, focused on the globular cluster NGC 7006, is part of the long effort to employ the RR Lyrae stars light curves as indicators of their physical parameters, via the Fourier decomposition. A recent summary of the results for 35 globular clusters can be found in paper by \citet{Arellano2022}. In that work the author argues in favour of the accuracy of the metallicities and distances rendered by the approach.

NGC 7006 is located in the constellation Delphinus ($\alpha = 21^{\rm h} 01^{\rm m} 29.38^{\rm s}$, $\delta = +16^{\circ} 11^{\prime} 14.4^{\prime \prime}$, J2000). Its present galactic position is  $l=63.77^{\circ}$,   $b=-19.41^{\circ}$ and Z=--13.7 kpc, although its long-term orbit show diversions as large as |Z|=56.13 kpc \citep{Allen2006} . It is a cluster rich in RR Lyrae stars, 64 of them are recorded in the Catalogue of Variable Stars in Globular Clusters \citep{Clement2001}. The only other type of variables known in the cluster before the present work were two red giant star of the types SR8 (V19) and L (V54). A comprehensive study of the 76 variables presently known in NGC 7006 and their periods was published by  \citet{Wehlau1999} (hereinafter WSN99).

In this paper we report the results of a time-series \emph{VI} CCD photometry of the variable stars in the field of the cluster, discuss the stellar cluster membership, estimate relevant physical parameters of RR Lyrae stars via the Fourier decomposition of their light curves, particularly the cluster mean metallicity and distance, discuss the horizontal branch (HB) structure, and the time scales and mass of the precursor main sequences stars of the present RR Lyrae. We also performed a routinely search of previously undetected variables and report the discovery of ten new variables; two RRab, two RRc, one CWB (or BL Her) star and five semi regular or SR stars (see \S \ref{search}). We discuss the non cluster member status of several variables and have noticed the need to correct a few variable types. Notes on individual stars are to be found in Appendix A.

\section{Observations and reductions}

CCD $VI$ image of NGC 7006 were obtained during 11 nights over 4 years: 2011 October 5-6, 2012 August 23-25, 2013 between July 30 and August 27, and 2015 June 26-27. The observations were performed with the 2 m telescope at the Indian Astronomical Observatory (IAO) in Hanle, India. The detector used was a SITe ST-002 2Kx4K CCD  pixels with a scale of 0.296 arcsec/pix, translating to a field of view (FoV) of approximately 10.1 $\times$ 10.1 arcmin$^2$.

Table \ref{tab:Observation} contains the log of the observations and includes the date, number of images and exposure time by filter and the mean seeing nightly conditions.

\begin{table}
    \begin{center}
    \caption{The distribution of observation of NGC 7006$^a$. }
    \begin{tabular}{cccccc} 
    \hline
    Date & N$_V$ & t$_V$(s) & N$_I$ & t$_I$(s) & Mean seeing ($^{\prime \prime}$) \\
    \hline
    2011/10/05 & 8 & 120-300 & 10 & 20-80 & 2.0 \\
    2011/10/06 & 7 & 100 & 8 & 40 & 1.9 \\
    2012/08/23 & 9 & 60-240 & 0 & - & 1.8 \\
    2012/08/25 & 55 & 60-240 & 0 & - & 1.8 \\
    2013/07/30 & 2 & 110-120 & 5 & 40-50 & 1.6 \\
    2013/07/31 & 22 & 105-110 & 26 & 40-55 & 1.9 \\
    2013/08/01 & 31 & 75-120 & 31 & 35-60 & 1.7 \\
    2013/08/26 & 28 & 100-110 & 27 & 50-60 & 1.7 \\
    2013/08/27 & 25 & 60-90 & 28 & 35-50 & 1.6 \\
    2015/06/26 & 4 & 300 & 6 & 130 & 2.0 \\
    2015/06/27 & 8 & 300 & 8 & 130 & 1.9 \\
    \hline
    Total: & 199 & & 149 & & \\
    \hline
    \multicolumn{6}{p{8cm}}{\footnotesize{$^a$Columns N$_V$ and N$_I$ indicate the number of images obtained in the $V$-band and $I$-band, respectively. Columns t$_V$ and t$_I$ are the range of exposure time used during each night. In the last column, the mean seeing in arc seconds for each night are listed.}}
    \end{tabular}
    \label{tab:Observation}
    \end{center}
\end{table}

\subsection{Difference image analysis}

All cluster frames were corrected via numerous bias and sky flat field images through standard procedures. For the extraction of accurate photometry, we employed the difference image analysis (DIA) technique, for which the DanDIA pipeline was used \citep{Bramich2008,Bramich2013}. 
This approach has been repeatedly used and described in detail in previous publications. The interested reader can find a thorough description of the approach and the conversion of fluxes into magnitudes in the  paper by \citet{Bramich2011}.

\subsection{Transformation to the standard system}

The transformation of the instrumental \emph{vi} magnitudes to the standard \emph{VI} Johnson–
Kron–Cousins standard system (Landolt 1992), was performed
using about 300 standard stars in the FoV of NGC 7006 included
in the catalogue of standard stars in globular clusters of \citep{Stetson2000}. Linear fits to the correlations $(v-i)$ vs. $(V-v)$ and $(v-i)$ vs. $(I-i)$,  carry a small colour dependence, as is generally expected. The transformation equations are form:

\begin{equation}
V-v = +0.044 (\pm 0.004) (v-i) - 1.096 (\pm 0.004), 
\label{transV}
\end{equation}

\begin{equation}
I-i = +0.031 (\pm 0.004) (v-i) - 1.314 (\pm 0.004), 
\label{transI}
\end{equation}

\begin{table}
\begin{center}
\caption{Time-series \emph{VI} photometry for the variables stars observed in this work$^a$.}
\begin{tabular}{cccccc}
\hline
Variable & Filter & HJD & $M_{\rm std}$ & $m_{\rm ins}$ & $\sigma_m$  \\
Star ID & & (d) & (mag) & (mag) & (mag)\\
\hline
V1 & V & 2455840.08949 & 19.201 & 20.286 & 0.022 \\
V1 & V & 2455840.09717 & 19.232 & 20.317 & 0.025 \\
\vdots & \vdots & \vdots & \vdots & \vdots & \vdots \\
V1 & I & 2455840.06083 & 18.479 & 19.785 & 0.055 \\
V1 & I & 2455840.08418 & 18.536 & 19.842 & 0.031 \\
\vdots & \vdots & \vdots & \vdots & \vdots & \vdots  \\
V2 & V & 2455840.08949 & 19.157 & 20.239 & 0.020 \\
V2 & V & 2455840.09717 & 19.168 & 20.250 & 0.022 \\
\vdots & \vdots & \vdots & \vdots & \vdots & \vdots \\
V2 & I & 2455840.06083 & 18.386 & 19.690 & 0.049 \\
V2 & I & 2455840.08418 & 18.451 & 19.755 & 0.027 \\
\vdots & \vdots & \vdots & \vdots & \vdots & \vdots \\
\hline
\multicolumn{6}{p{8cm}}{\footnotesize{$^a$The variable star, filter used and epoch of mid-exposure are listed in columns 1, 2 and 3, respectively. The instrumental magnitudes, $m_{ins}$, and standard magnitudes, $m_{std}$, corresponding to each variable star are listed in the columns 4 and 5, in addition, the column 6 contains the uncertainty of the instrumental magnitude, $\sigma_m$, which also corresponds to the error of the standard magnitude. The full version of this table is available at the CDS data base.}}\\
\end{tabular}
\label{tab:Photometry}
\end{center}
\end{table}

In Table \ref{tab:Photometry}, we show a small fragment of the resulting time-series $VI$ photometry for each variable star obtained in this work. The full table shall be available in electronic form in the {\it Centre de Donnes astronomique de Strasbourg} data base (CDS).

\section{Star Membership in the FoV of NGC 7006} 
\label{MEMBERSHIP}

To establish the membership status of the stars in our FoV, we have employed the high-quality astrometric data available in {\it Gaia}-DR3 \citep{Gaia2021} and the method developed by \cite{Bustos2019}, which consists of two main stages: 
\begin{itemize}
    \item The first stage is based on a clustering algorithm applied to a multidimensional space of physical parameters. It aims to find groups of stars that possess similar characteristics in the four-dimensional space of the gnomic coordinates ($X_{\rm t}$,$Y_{\rm t}$) and proper motions ($\mu_{\alpha*}$,$\mu_\delta$).
    \item The second stage consists of  the analysis of the projected distribution of stars with different proper motions around of the mean proper motion of the cluster, and it is aimed to identify probable missing members.
\end{itemize}

\begin{figure*}
    \centering
    \includegraphics[width=0.9\textwidth]{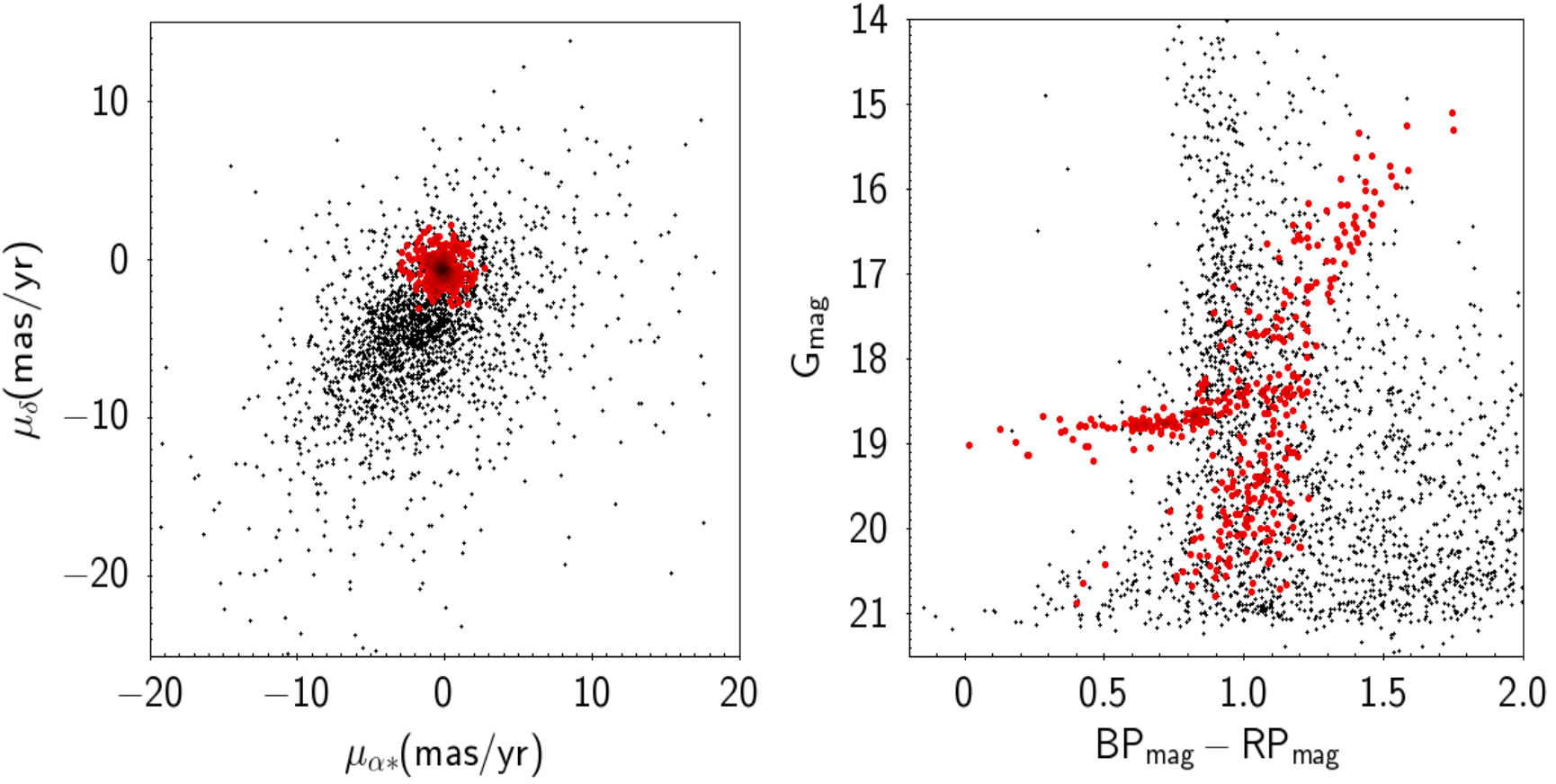}
    \caption{Red and gray points correspond to the stars that were found cluster members and non-members, respectively. Left-hand panel: VPD of the cluster NGC 7006; right-hand panel: color-magnitude diagram of NGC 7006 in the Gaia's photometric system.}
    \label{fig:VPD}
\end{figure*}

The method was applied to 3316 Gaia-sources in a field of radius 10 arcmin centered in the cluster, of which 530 were identified as likely cluster members. Fig. \ref{fig:VPD} displays the Vector Point Diagram (VPD), where the distribution of the proper motions is shown, and the resulting Color-Magnitude diagram (CMD) of likely cluster members.

We were able to measure light curves for 1373 stars in the FoV of our images, 470 of which are among the likely cluster members, enabling to refine the working CMD that shall be discussed later in the paper. 

\section{Variable stars in NGC 7006}

\subsection{Search for new variables}
\label{search}

Taking advantage of the resulting time-series \emph{VI} for the 470 stars likely to be cluster members in out FoV, we performed a routine search for new variables. For this purpose
we splitted the CMD of NGC 7006 into regions where variable stars are typically found.
We selected four regions: the Blue Stragglers (BS), the Horizontal Branch (HB), the W Virginis and BL Herculis region above of the HB, and the Tip of the Red Giant Branch (TRGB). We analyzed the light curves of the stars belonging to these regions, and calculated their period using the string-length method \citep{Burke1970, Dworetsky1983}, and phase their light curves.  With this method we discovered 10 new variables and have assigned them variable numbers as follows: two RRc (V77 and V78), two RRab stars (V79 and V80), one BL Hercules star (V81) and five SR stars (V82, V83, V84, V85 and V86). Their folded light curves are shown in Fig. \ref{fig:RRL1}.

\cite{Gerashchenko2006} had suggested the existence of two new variable stars which he called V77 and V78, however we identified these stars in our photometry collection and found them not to be variable. These stars are not registered in the CVSGC, so we do not take them into account when naming the stars discovered in this work.

\begin{figure*}
\centering
\includegraphics[width=18cm,height=9.0cm]{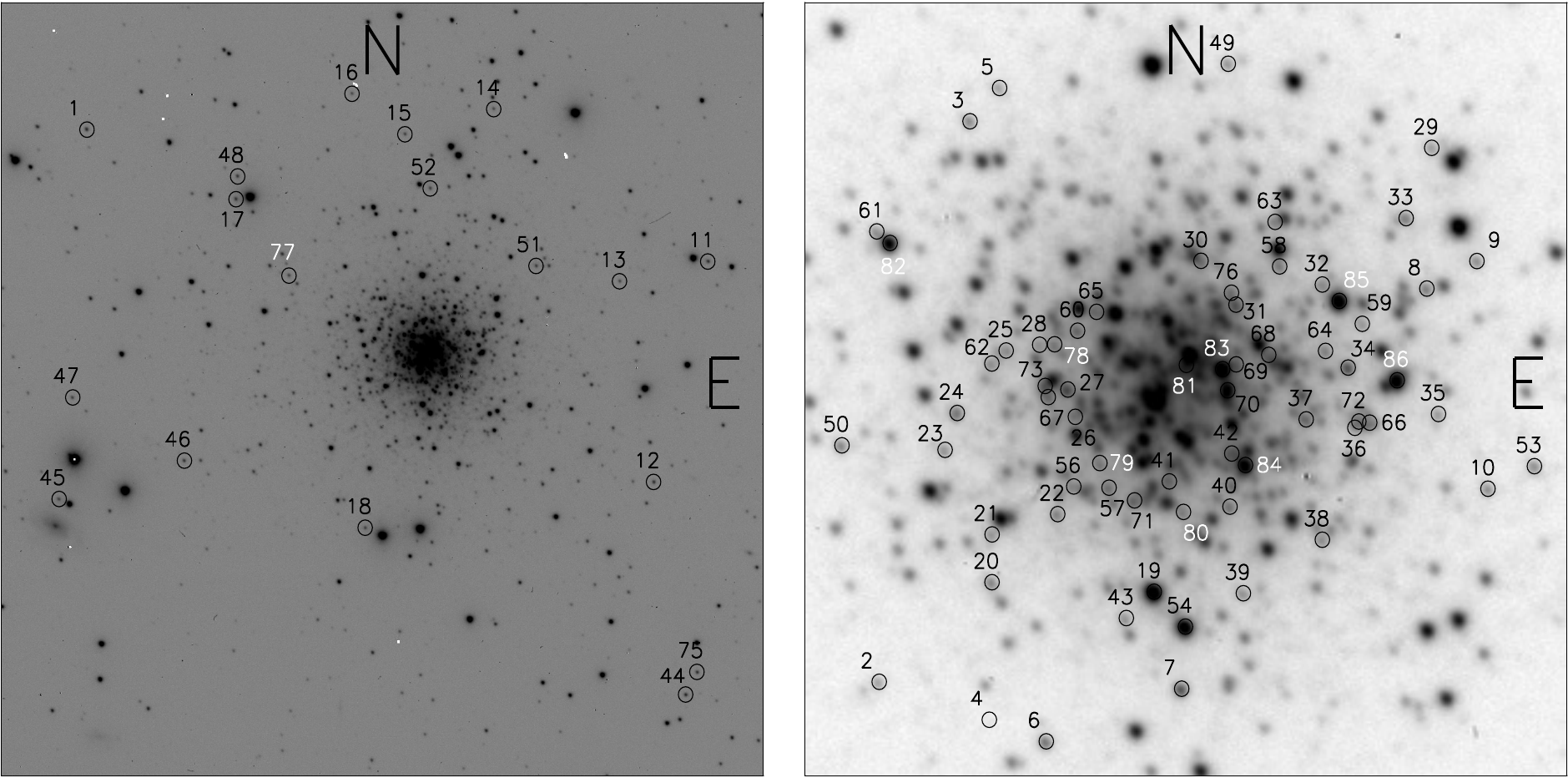}
\caption{Identification charts of all variable stars in NGC 7006 listed in Table \ref{tab:periods}. The panels are constructed from the $V$ reference image, built by stacking the best quality images in our collection. The left panel shows a field of 6.7 $\times$ 6.7 arcmin$^2$. The panel to the right displays the central region of the cluster with a field of  $1.63 \times 1.63$ arcmin$^2$. The labels refer to the variable numbers and those labeled in white are newly discovered in this work.}
\label{fig:astrometry}
\end{figure*}

Table \ref{tab:periods} lists all the variable stars presently known in NGC 7006 and shows the basic photometric and positional data and membership status for each variable. The reported periods found in our data are listed in column 7, while in column 8 we include the periods reported by WSN99. With a few exceptions the periods are very similar. All variable stars listed in Table \ref{tab:periods} have been identified in the charts of Fig. \ref{fig:astrometry}.

\begin{table*}
\caption{Data of variable stars in NGC 7006 in the FoV of our images.}
\centering
\begin{tabular}{clcccccccccc}
\hline
Variable & Variable & $\langle V\rangle$ & $\langle I\rangle$ &	$A_V$ & $A_I$ & $P$(d) & $P$(d) & HJD$_{max}$ & RA & Dec & Mem$^b$ \\
Star ID & Type & (mag) & (mag) & (mag) & (mag) & This work & WSN99 & $(+2 450 000)$ & (J2000.0) & (J2000.0) & \\ 
\hline
V1 & RRab & 18.901 & 18.389 & 1.121 & 0.794 & 0.492730 & 0.492731 & 6531.1366 & 21:01:16.81 & +16:13:07.1 & Y \\
V2 & RRab & 18.916 & 18.329 & 0.961 & 0.578 & 0.586961 & 0.586984 & 6505.3307 & 21:01:26.73 & +16:10:34.9 & Y \\
V3 & RRab Bl & 18.943 & 18.292 & 0.806 & 0.420 & 0.559646 & 0.560555 & 6163.2429 & 21:01:27.48 & +16:11:46.2 & Y \\
V5 & RRab & 18.839 & 18.291 & 0.832 & 0.607 & 0.535095 & 0.533288 & 6532.2167 & 21:01:27.74 & +16:11:50.4 & Y \\
V6 & RRab & 19.005 & 18.427 & 1.109 & 0.704 & 0.498056 & 0.498032 & 6532.1130 & 21:01:28.23 & +16:10:27.6 & Y \\
V8 & RRab & 19.097 & 18.384 & 0.913 & 0.651 & 0.564299 & 0.564293 & 6165.2675 & 21:01:31.58 & +16:11:25.6 & Y \\
V10 & RRab & 18.942 & 18.387 & 0.920 & 0.694 & 0.542048 & 0.542908 & 7201.3771 & 21:01:32.15 & +16:11:00.3 & Y \\
V11 & RRab & 18.958 & 18.322 & 0.864 & 0.505 & 0.576019 & 0.576032 & 6532.1296 & 21:01:39.53 & +16:12:02.4 & Y \\
V12 & RRab Bl & 18.902 & 18.259 & 0.755 & 0.441 & 0.574076 & 0.576032 & 6532.1678 & 21:01:37.64 & +16:10:08.0 & Y \\ 
V13 & RRab & 18.953 & 18.364 & 0.990 & 0.667 & 0.551647 & 0.551646 & 6531.1366 & 21:01:36.31 & +16:11:51.7 & Y \\
V14 & RRab & 18.958 & 18.327 & 0.917 & 0.641 & 0.560362 & 0.560360 & 6506.4312 & 21:01:31.65 & +16:13:20.1 & Y \\
V15 & RRab & 18.883	& 18.239 & 0.852 & 0.599 & 0.588070 & 0.588070 & 6506.3946 & 21:01:28.41 & +16:13:06.5 & Y \\
V16	& RRab Bl & 18.858 & 18.214 & 1.170 & 0.789 & 0.537565 & 0.537576 & 6532.1760 & 21:01:26.47 & +16:13:27.2 & Y \\
V17 & RRab & 18.934 & 18.389 & 1.104 & 0.869 & 0.511497 & 0.511497 & 6506.4556 & 21:01:22.28 & +16:12:31.9 & Y \\
V18 & RRab Bl & 18.955 & 18.314 & 0.764 & 0.551 & 0.603724 & 0.603707 & 6505.2748 & 21:01:27.13 & +16:09:42.6 & Y \\
V19	& SR & 15.601 & 14.047 & - & - & 41.0 & 92.17 & 6504.4353 & 21:01:29.18 & +16:10:46.7 & Y \\
V20 & RRab & 18.961 & 18.381 & 0.861 & 0.643 & 0.577536 & 0.577533 & 6165.3221 & 21:01:27.73 & +16:10:47.7 & Y \\
V21 & RRab & 18.910 & 18.236 & 0.786 & 0.459 & 0.612548 & 0.612559 & 6165.3559 & 21:01:27.72 & +16:10:53.8 & Y \\
V22 & RRab & 18.935 & 18.367 & 1.048 & 0.633 & 0.526925 & 0.526925 & 6165.2536 & 21:01:28.31 & +16:10:56.4 & Y \\
V23 & RRab & 18.797 & 18.193 & 0.876 & 0.561 & 0.606899 & 0.607949 & 6163.2429 & 21:01:27.29 & +16:11:04.4 & Y \\
V24 & RRab & 18.356 & 17.918 & 0.396 & 0.301 & 0.609384 & 0.627156 & 6163.2513 & 21:01:27.40 & +16:11:09.1 & N \\
V25 & RRab Bl & 18.844 & 18.392 & 0.913 & 0.735 & 0.543231 & 0.543232 & 6532.1678 & 21:01:27.83 & +16:11:17.1 & Y \\
V26 & RRab & 18.628$^a$ & - & 0.9$^a$ & - & - & 0.540697 & - & 21:01:28.44$^a$ & +16:11:08.6$^a$ & ? \\
V27	& RRab & 18.257 & 17.737 & 0.582 & 0.371 & 0.564680 & 0.564124 & 6504.4440 & 21:01:28.38 & +16:11:12.3 & N \\
V28 & RRd? & 19.012 & 18.465 & 1.191 & 0.748 & 0.33213 & 0.496958 & 6531.2216 & 21:01:28.13 & +16:11:17.9 & Y \\
&  & &  & &  & 0.49914 &   &   &   &   &   \\
V29 & RRab & 18.975 & 18.350 & 0.976 & 0.717 & 0.559245 & 0.559197 & 6504.4353 & 21:01:31.6 & +16:11:43.4 & Y \\
V30 & RRab & 18.634 & 18.221 & 0.661 & 0.571 & 0.568951 & - & 6506.3704 & 21:01:29.56 & +16:11:28.8 & ? \\
V31 & RRab & 18.901 & 18.092 & 0.966 & 0.497 & 0.563464 & 0.563139 & 6532.1110 & 21:01:29.88 & +16:11:23.3 & Y \\
V32 & RRab & 18.756 & 18.214 & 0.733 & 0.503 & 0.562648 & 0.562653 & 6506.3921 & 21:01:30.65 & +16:11:26.0 & Y \\
V33 & RRab & 19.032 & 18.348 & 0.997 & 0.625 & 0.556801 & 0.556807 & 6532.1296 & 21:01:31.39 & +16:11:34.5 & Y \\
V34	&EW? & 17.884 & 16.744 & - & - & 0.908797 & - & 6163.2557 & 21:01:30.89 & +16:11:15.5 & Y \\
V35	& RRab & 18.983 & 18.313 & 0.727 & 0.564 & 0.596268 & 0.596265 & 6163.2578 & 21:01:31.70 & +16:11:09.7 & Y \\
V36 & RRc & - & - & - & - & - & 0.276970 & - & 21:01:30.94$^a$ & +16:11:07.6$^a$ & ? \\
V37 & RRab & - & - & - & - & - & 0.5679$^d$ & - & 21:01:30.50$^a$ & +16:11:08.1$^a$ & Y \\
V38 & RRab & 18.837 & 18.282 & 0.830 & 0.644 & 0.622316 & 0.624434 & 6531.2191 & 21:01:30.68 & +16:10:53.6 & Y \\
V39 & RRab & - & - & - & - & - & 0.578450 & - & 21:01:29.99$^a$ & +16:10:46.4$^a$ & Y \\
V41 & RRab & 18.635 & 17.914 & 0.762 & 0.427 & 0.566347 & 0.566852 & 6165.3736 & 21:01:29.30 & +16:11:00.8 & N \\
V42 & RRab & 18.458 & 17.941 & - & - & 0.746606 & - & 6165.3559 & 21:01:29.86 & +16:11:04.4 & N \\
V43 & RRab & 18.944 & 18.330 & 0.622 & 0.471 & 0.596600 & 0.596614 & 6531.2216 & 21:01:28.93 & +16:10:43.4 & Y \\
V44 & RRab & 18.910 & 18.285 & 0.727 & 0.481 & 0.587795 & 0.587779 & 6506.2512 & 21:01:38.90 & +16:08:18.1 & Y \\
V45 & RRab & 18.903 & 18.229 & 0.752 & 0.504 & 0.583846 & 0.583852 & 6531.2897 & 21:01:15.95 & +16:09:55.7 & Y \\
V46 & RRab & 18.872 & 18.272 & 0.318 & 0.234 & 0.668906 & 0.667587 & 6532.1130 & 21:01:20.51 & +16:10:16.3 & Y \\
V47 & RRab & 18.949 & 18.343 & 0.653 & 0.515 & 0.567464 & 0.568285 & 6532.1739 & 21:01:16.41 & +16:10:48.3 & Y \\
V48	& RRab & 18.967 & 18.304 & 0.718 & 0.522 & 0.611972 & 0.611975 & 6506.4556 & 21:01:22.33 & +16:12:43.7 & Y \\
V49 & RRab & 18.984 & 18.355 & 0.859 & 0.434 & 0.581855 & 0.581895 & 6506.3489 & 21:01:29.78 & +16:11:53.8 & Y \\
V50	& RRab & 18.919 & 18.317 & 0.835 & 0.630 & 0.590330 & 0.590383 & 6505.4599 & 21:01:26.37 & +16:11:04.9 & Y \\
V51	& RRab & 18.977 & 18.254 & 0.439 & 0.315 & 0.643829 & 0.643838 & 6505.4599 & 21:01:33.26 & +16:11:59.1 & Y \\
V52 & RRab & 18.938 & 18.246 & 0.550 & 0.465 & 0.621717 & 0.621703 & 6506.2459 & 21:01:29.36 & +16:12:38.7 & Y \\
V53 & RRc & 18.837 & 18.500 & 0.248 & 0.230 & 0.260171 & 0.357694 & 6165.2674 & 21:01:32.56 & +16:11:03.2 & Y \\
V54 & L & 16.081 & 14.778 & - & - & 30.7 & - & 6163.2513 & 21:01:29.46 & +16:10:42.4 & Y \\
V55 & RRab & - & - & - & - & - & 0.537738 & - & 21:01:12.40$^a$ & +16:16:13.9$^a$ & ? \\ 
V56 & RRab & 19.086 & 18.365 & 1.097 & 0.713 & 0.547423 & 0.549655 & 6531.1929 & 21:01:28.45 & +16:10:59.9 & Y \\
V57 & RRc & 18.956 & 18.518 & 0.630 & 0.407 & 0.351903 & 0.351893 & 6532.1657 & 21:01:28.76 & +16:10:59.9 & Y \\
V58 & RRab & 18.851 & 18.157 & 0.256 & 0.213 & 0.743666 & 0.5150$^d$ & 6532.1110 & 21:01:30.27 & +16:11:28.2 & Y \\
V59	& RRab & 18.975 & 18.463 & 1.093 & 0.830 & 0.480997 & 0.480985 & 5841.0670 & 21:01:31.01 & +16:11:21.0 & Y \\
V60 & RRab & 18.913 & 18.189 & 0.611 & 0.385 & 0.626593 & 0.385810 & 6506.3704 & 21:01:28.46 & +16:11:19.7 & Y \\
V61 & RRab & 18.816 & 18.284 & 1.025 & 0.645 & 0.589081 & 0.589097 & 6506.2858 & 21:01:26.66 & +16:11:32.1 & N \\
V62 & RRc & 18.873 & 18.349 & 0.439 & 0.322 & 0.346931 & 0.346946 & 6532.1981 & 21:01:27.70 & +16:11:15.5 & Y \\
V63 & RRab Bl & 18.800 & 18.296 & 0.957 & 0.699 & 0.547209 & 0.5280$^d$ & 5840.1197 & 21:01:30.22 & +16:11:33.8 & Y \\
\hline
\end{tabular}
\label{tab:periods}
\end{table*}

\begin{table*}
\addtocounter{table}{-1}
\caption{Continued }
\centering
\begin{tabular}{clcccccccccc}
\hline
Variable & Variable & $\langle V\rangle$ & $\langle I\rangle$ &	$A_V$ & $A_I$ & $P$(d) & $P$(d) & HJD$_{max}$ & RA & Dec & Mem$^b$ \\
Star ID & Type & (mag) & (mag) & (mag) & (mag) & This work & WSN99 & $(+2 450 000)$ & (J2000.0) & (J2000.0) & \\
\hline
V64 & RRc & 18.877 & 18.528 & 0.386 & 0.296 & 0.306375 & 0.313142 & 6504.4353 & 21:01:30.68 & +16:11:17.5 & Y \\
V65 & RRab & 18.776 & 18.220 & 0.563 & 0.456 & 0.558761 & - & 6165.3559 & 21:01:28.63 & +16:11:22.2 & Y \\
V66 & RRc & 18.452 & 17.628 & 0.339 & 0.141 &0.361055 & 0.6172$^d$ & 6506.3161 & 21:01:31.09 & +16:11:08.5 & Y \\
V67 & RRab & 18.508 & 17.699 & 0.408 & 0.213 & 0.554945 & 0.3175$^d$ & 6504.4440 & 21:01:28.21 & +16:11:11.3 & N \\
V68 & RRab & 19.015 & 18.323 & 0.995 & 0.641 & 0.555726 & - & 6506.3562 & 21:01:30.18 & +16:11:16.9 & Y \\
V69 & RRab & 19.422 & 18.615 & 1.466 & 0.727 & 0.656633 & - & 5841.0670 & 21:01:29.89 & +16:11:15.7 & N \\
V70 & CW & 16.680 & 15.570 & - & - & 11.7 & - & 6165.3559 & 21:01:29.81 & +16:11:12.4 & Y \\
V71 & RRc & 18.703 & 18.198 & 0.494 & 0.306 & 0.335231 & 0.348346 & 6165.3559 & 21:01:28.99 & +16:10:58.3 & N \\
V72 & RRc & - & - & - & - & - & - & - & 21:01:30.96$^a$ & +16:11:09.8$^a$ & ? \\
V73 & RRab & 18.129	& 17.276 & 0.566 & 0.341 & 0.554953 & 0.5780$^d$ & 6165.3580 & 21:01:28.18 & +16:11:12.7 & N \\
V74 & = V41  \\
V75 & CWA & 18.628 & 17.917 & 0.121 & 0.086 & 13.5071 & - & 6506.4556 & 21:01:39.30 & +16:08:29.9 & N \\
V76 & RRab & 19.118 & 18.195 & 1.240 & 0.598 & 0.566262 & 0.56170 & 6532.1657 & 21:01:29.84 & +16:11:24.8 & Y? \\
V77$^c$ & RRc & 19.044 & 18.252 & 0.086 & 0.074 & 0.317159 & - & 6165.3600 & 21:01:24.24 & +16:11:52.7 & Y \\
V78$^c$ & RRc & 18.805 & 18.176 & 0.261 & 0.174 & 0.332177 & - & 6505.3514 & 21:01:28.26 & +16:11:18.0 & Y \\
V79$^c$ & RRab & 19.055 & 18.363 & 0.804 & 0.653 & 0.598871 & - & 6531.3100 & 21:01:28.68 & +16:11:03.0 & Y \\
V80$^c$ & RRab & 19.181 &	18.809 & 0.119 & 0.098 & 0.541755 & - & 6165.2563 & 21:01:29.43 & +16:10:56.9 & Y \\
V81$^c$ & CWB & 17.888 & 17.185 & - & - & 4.63 & - & 6165.3559 & 21:01:29.44 & +16:11:15.6 & Y \\	
V82$^c$ & SR & 16.455 & 15.131 & - & - & 14.01 & - & 6504.4440 & 21:01:26.78 & +16:11:30.6 & Y \\
V83$^c$ & SR & 15.934 & 14.278 & - & - & 18.68 & - & 6532.1110 & 21:01:29.77 & +16:11:15.1 & Y \\
V84$^c$ & SR & 16.266 & 14.940 & - & - & 10.22 & - & 6165.3538 & 21:01:29.98 & +16:11:02.9 & Y \\	
V85$^c$ & SR & 15.858 & 14.262 & - & - & 8.37 & - & 6165.3642 & 21:01:30.80 & +16:11:23.9 & Y \\
V86$^c$ & SR & 16.215 & 14.775 & - & - & 36.14 & - & 6532.1110 & 21:01:31.33 & +16:11:13.9 & Y \\
\hline
\multicolumn{12}{l}{Bl: RR Lyrae with Blazhko effect. Amplitudes for this star correspond to the maximum observed.} \\
\multicolumn{12}{l}{$^a$ Value taken from CVSGC \citep{Clement2001}.} \\
\multicolumn{12}{l}{$^b$ Membership status: Y = member, N = no-member, ? no proper motions available, or dubious membership.} \\
\multicolumn{12}{l}{$^c$ New variable star found in this work.}\\
\multicolumn{12}{l}{$^d$ Period from \citet{PintoRosin1973}}\\
\end{tabular}
\end{table*}

\begin{figure*}
\centerline{\includegraphics[width=16cm]{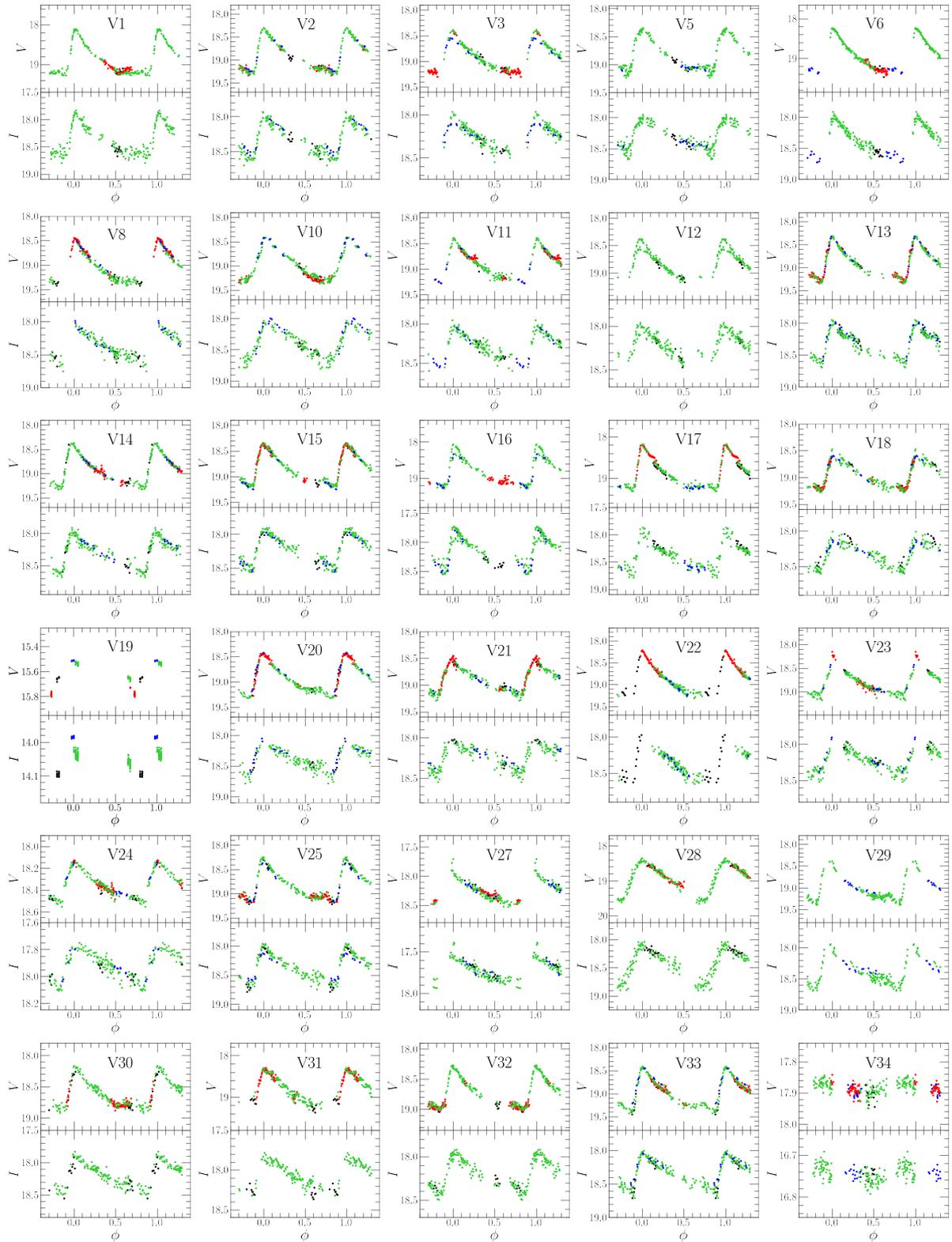}}
\caption{Light curves of 75 variable in the field of view of NGC 7006. For membership conclusions refer to the last column of Table \ref{tab:periods}. All light curves are phased with the ephemerides given in that table. V75 is plotted as function of HJD to highlight its long-term variation. For a phased light curve and a discussion of this star please refer to AppendixA.  In this mosaic the symbols colours are coded as: black, red, green and blue for the 2011, 2012, 2013 and 2015 seasons.}
\label{fig:RRL1}
\end{figure*} 

\begin{figure*}
\centerline{\includegraphics[width=16cm]{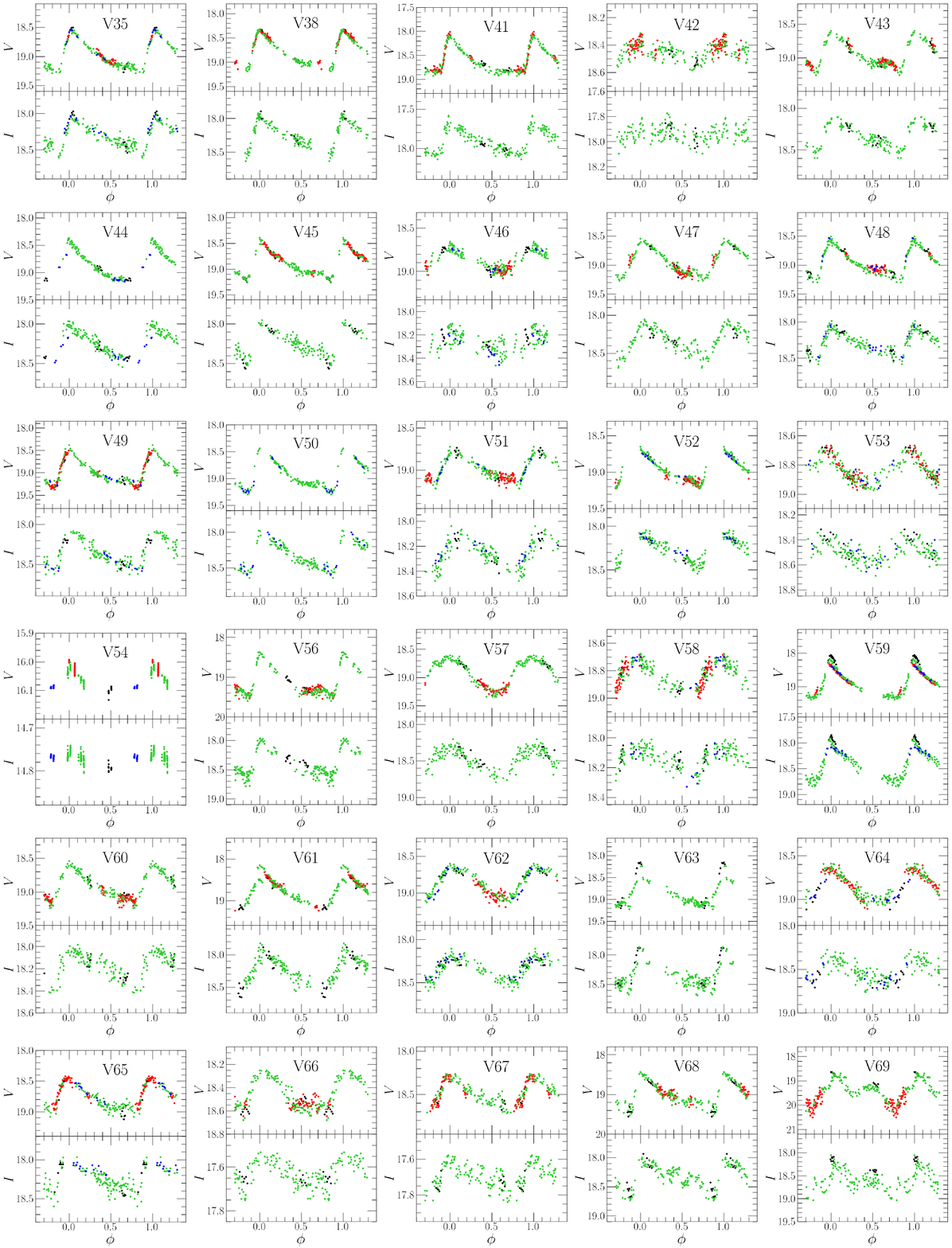}}
\addtocounter{figure}{-1}
\caption{continued}
\label{fig:RRL2}
\end{figure*} 

\begin{figure*}
\centerline{\includegraphics[width=16cm,height=20cm]{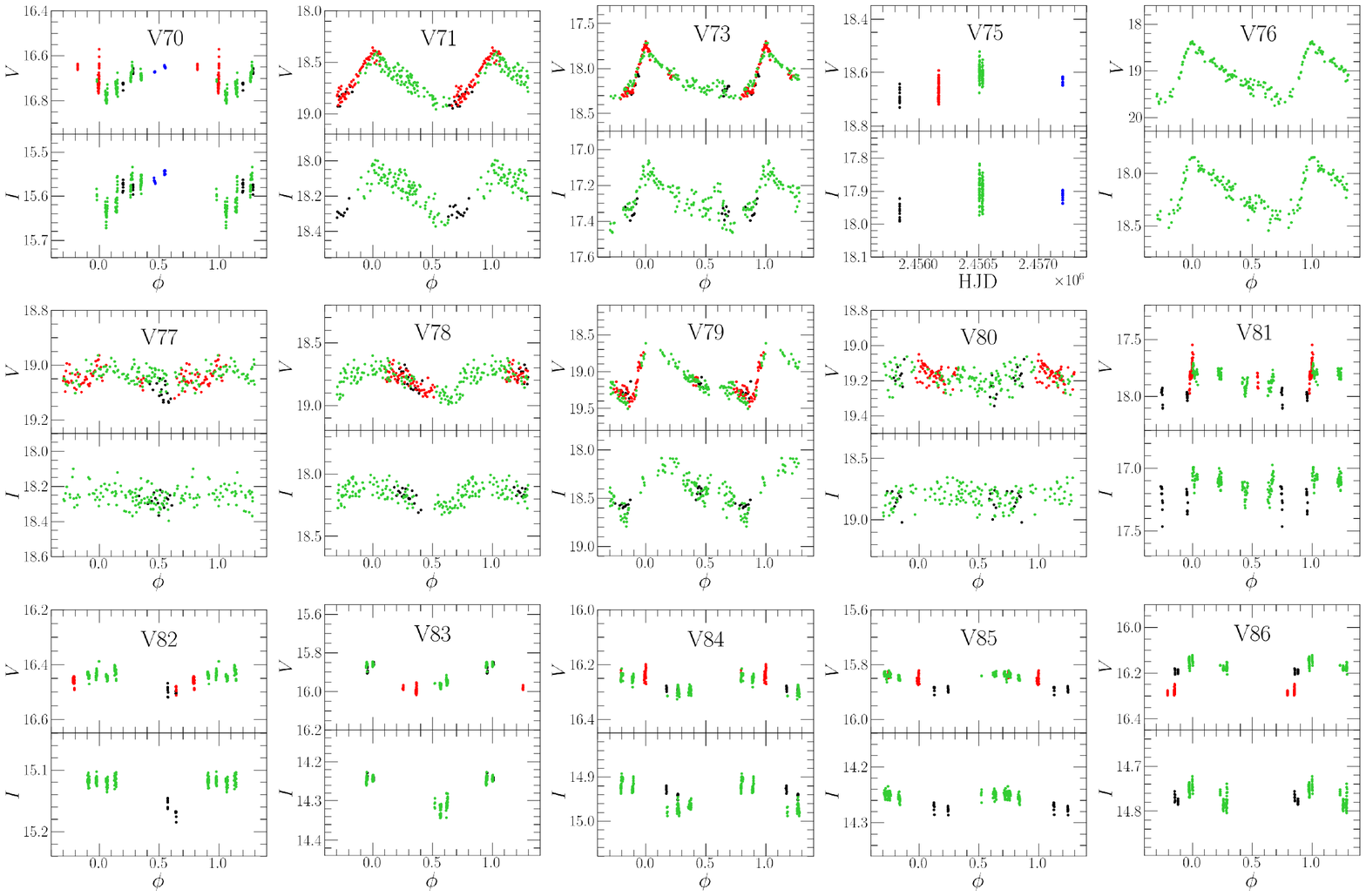}}
\addtocounter{figure}{-1}
\caption{continued}
\label{fig:RRL3}
\end{figure*}

\subsection{The RR Lyrae stars}

\subsubsection{RR Lyrae variable stars in NGC 7006}

During the analysis we noticed that several stars had a wrong classification or coordinates slightly off the position of the corresponding variable both in our astrometrically calibrated image and in the $GAIA$-DR3 release. The corrected types and coordinates are those listed in Table \ref{tab:periods}.

The total number of RR Lyrae stars in Table \ref{tab:periods} is 71 but a few stars were considered field stars, as indicated in the last column of this table. There are 49 RRab, 7 RRc, 1 RRd likely cluster members.  Furthermore, we noticed that the RRab stars V3, V12, V16 and V63 show Blazhko-like amplitude modulations not previously reported. Please refer to Appendix A for details on individual stars.

\subsubsection{Oosterhoff type}

The average period of the RRab stars in NGC 7006 is $<P_{ab}> = 0.577 \pm 0.006$ d, the estimated uncertainty is the standard error of the mean, and the fraction of RRc stars is $N_c/(N_c + N_{ab}) = 0.14$. These values clearly identify  NGC 7006 as a Oosterhoff type OoI, consistent with the classification made by \citet{Gerashchenko2018}. 

\begin{figure}
\centering
\includegraphics[width=1\columnwidth]{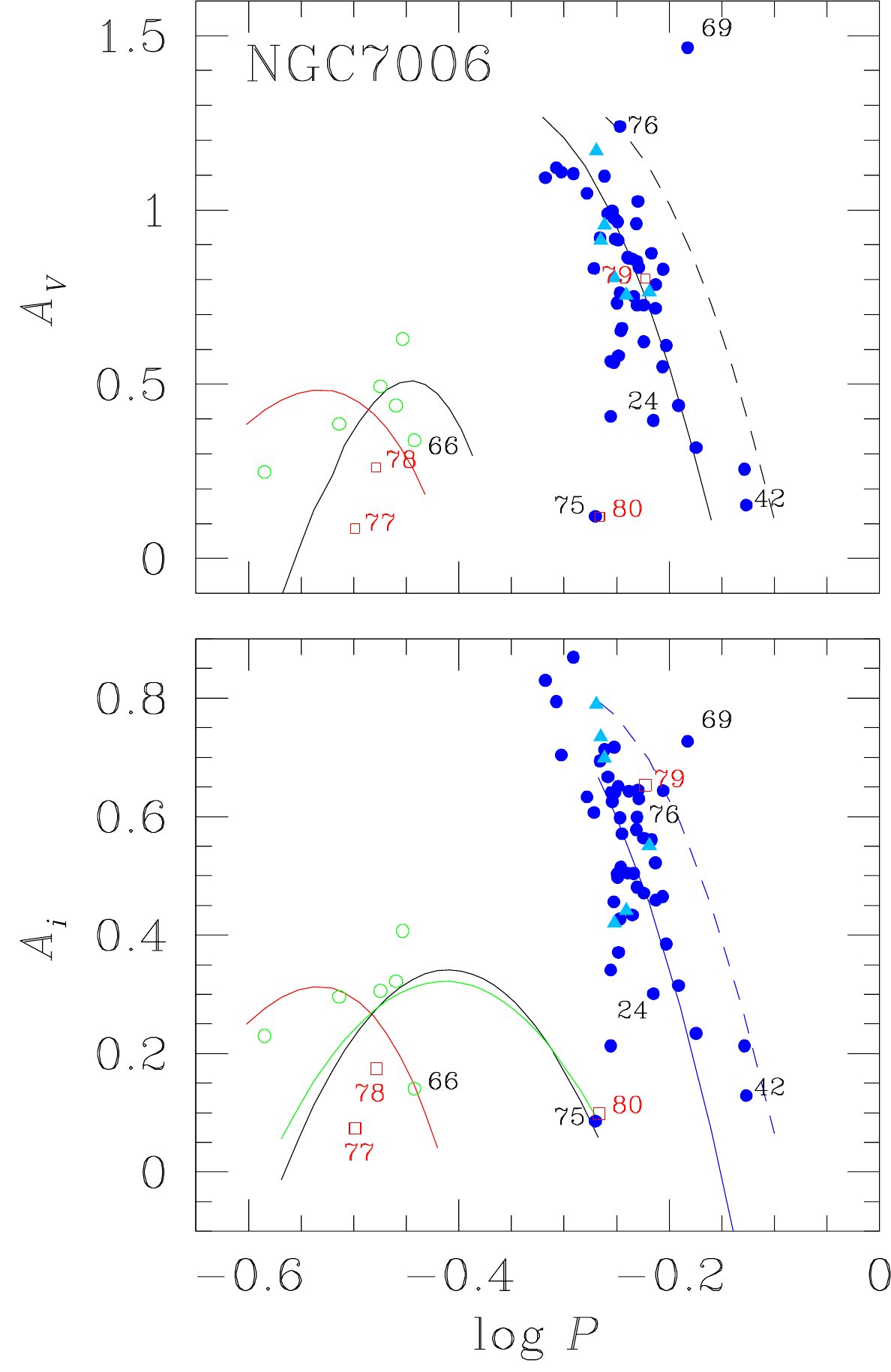}
\caption{Period-Amplitude plane in $V$ (top) and $I$ (bottom) filters for NGC 7006. Blue and green symbols represent RRab and RRc stars respectively. The stars that present Blazhko-like amplitude modulations are marked with light blue triangles. The empty red squares indicate the new variables announced in the present work. In the top panel, the continuous and segmented black loci represent the distribution of unevolved and evolved stars, respectively, in M3 according to \citep{Cacciari2005}. \citet{Kunder2013b} found the black parabola for the RRc stars from 14 OoII clusters. \citet{Arellano2015} calculated the red parabolas from a sample of RRc stars in five OoI clusters, excluding variables with Blazhko effect. In the bottom panel, the continues and segmented blue lines were constructed by \citet{Kunder2013a}. The green parabola was calculated by \citet{Deras2019} from RRc stars in M13 and the black one by \citet{Yepez2020}, using 28 RRc stars from seven OoII clusters.}
\label{fig:Bailey}
\end{figure}

Furthermore, the period-amplitude plane for RR Lyrae stars, also known as the Bailey diagram shown in Fig. \ref{fig:Bailey} for the $V$ and $I$ light curve amplitudes, also confirm the OoI type of the cluster. The amplitudes were estimated
from the best fit obtained by the Fourier decomposition of the light curves. For the stars with Blazhko-like modulations, the maximum amplitude was taken. The distribution of RRab stars and the reference loci for unevolved and evolved stars (OoI and OoII), described in the figure caption, clearly confirm the cluster as of the OoI type. The anomalous position of the some stars in the Bailey diagram will be discussed in Appendix A.

\subsubsection{Reddening estimate from RRab stars}

\citet{Sturch1966}, demonstrated that in RRab stars the minimum colour between phases 0.5 and 0.8, $(V-I)_{min}$, is constant. \citet{Guldenschuh2005} calibrated the intrinsic color between the same phases as $(V-I)_{0,min} = 0.58 \pm 0.02$. Using these results, reddening $E(V-I)$ can be estimated. We obtained the value of $E(B-V)$ from the ratio $E(V-I)/E(B-V) = 1.259$ \citep[e.g.][]{Yepez2020}. Employing 41 RRab stars whose light curves were of best quality we obtained an average reddening $E(B-V) = 0.08 \pm 0.05$. The large uncertainty is mainly due to the noise in our light curves.  

It is of interest to recall the values of interstellar reddening estimated from the position of the cluster by \citet{Schlafly2011}, $E(B-V)_{S\&F} = 0.07 \pm 0.003$, and \citet{Schlegel1998} (SFD), $E(B-V)_{SFD} = 0.08 \pm 0.004$. Henceforth we shall adopt $E(B-V)=0.08$ for the rest of the present work.

\subsection{Other variable stars}
 While the variable star population of NGC 7006 is dominated by the presence of RR Lyrae stars, luminous variables near the TRGB are also present:
 seven SR type (V19, V70, V82, V83, V84, V85 and V86), one L type (V54) and one that we could not classify (V34). Their light curves are shown in figure \ref{fig:RRL1} phased with the periods listed in Table \ref{tab:periods} which are accurate only to the second digit.

\section{Physical parameters of RR Lyrae stars}

The light-curve morphology of RR Lyrae stars is tightly related to some of fundamental physical parameters of the star and of the cluster to which they belong, of particular relevance being the metallicity or [Fe/H] and the distance. The Fourier decomposition of the light curve in harmonics, is parameterized by an equation of the form 

\begin{equation}
m(t) = A_0 + \sum_{k=1}^{N} A_k \cos\left(\frac{2\pi k}{P}(t-E) + \phi_k\right),  
\label{eq:fourier}
\end{equation}

\noindent
where $m(t)$ is the magnitude at time $t$, $P$ is the period in days, $E$ is the epoch or time of maximum light and $N$ is the degree or number of harmonics used to reproduce the curve. $A_0$ is the intensity weighted mean, that is the best estimation of mean magnitude of the light curve. We used a linear minimization routine to derive the best fit values of the amplitude $A_k$ and phase $\phi_k$ of each harmonic and from which we can calculate the Fourier parameters defined as $\phi_{ij} = j\phi_i - i\phi_j$ and $R_{ij} = A_i/A_j$, with $1 \leq i, j \leq N$.

This approach to the calculation of the mean metallicity and distance of globular clusters with RR Lyrae stars has proven to be very accurate and the results compare very well, within the respective uncertainties with spectroscopic [Fe/H] values and distances obtained via Gaia-DR3 and HST, as it has been recently discussed by \citet{Arellano2022}  for a sample of 37 globualr clusters.

\begin{table*}
\caption{Fourier coefficients of the $V$ light curves of RRab and RRc stars in NGC 7006. The deviation parameter $D_m$ is also listed for the RRab stars.}
\centering
\begin{tabular}{cccccccccc}
\hline
Variable & $A_0$ & $A_1$ & $A_2$ &	$A_3$ & $A_4$ &	$\phi_{21}$ & $\phi_{31}$ & $\phi_{41}$ & $D_{\rm m}$ \\
ID & (V mag) & (V mag) & (V mag) & (V mag) & (V mag) & & & & \\
\hline
\multicolumn{10}{c}{RRab} \\
\hline
V1 & 18.901(4) & 0.406(5) &	0.182(5) & 0.146(5)	& 0.076(5) & 3.793(39) & 7.932(53) & 5.585(85) & 4.9 \\
V2 & 18.916(4) & 0.332(6) &	0.176(6) & 0.107(5) & 0.088(6) & 4.194(46) & 8.250(74) & 6.684(93) & 5.7 \\
V3 & 18.943(4) & 0.317(5) & 0.143(5) & 0.095(5) & 0.056(5) & 4.047(52) & 8.303(72) & 6.109(112) & 3.3 \\
V5 & 18.839(5) & 0.303(7) & 0.128(7) & 0.091(8) & 0.048(7) & 3.905(75) & 8.031(103) & 6.307(178) & 2.6 \\
V6 & 19.005(5) & 0.413(8) & 0.183(7) & 0.142(6) & 0.087(6) & 3.886(54) & 7.905(76) & 6.002(112) & 6.2 \\
V8 & 19.097(4) & 0.341(5) & 0.162(6) & 0.122(5) & 0.068(5) & 3.848(47) & 8.103(67) & 6.093(106) & 1.9 \\
V10 & 18.942(5) & 0.385(7) & 0.125(6) & 0.071(7) & 0.026(5) & 4.416(62) & 8.487(96) & 6.322(287) & 6.0 \\
V11 & 18.958(3) & 0.299(5) & 0.130(5) & 0.104(5) & 0.064(5) & 3.888(51) & 8.263(70) & 6.215(100) & 5.3 \\
V12 & 18.958(3) & 0.286(5) & 0.126(7) & 0.095(6) & 0.049(6) & 3.649(72) & 7.778(102) & 5.536(162) &2.0 \\
V13 & 18.953(5) & 0.324(7) & 0.159(7) & 0.113(7) & 0.082(7) & 3.757(60) & 7.997(83) & 5.815(117) & 2.7 \\
V14 & 18.958(4) & 0.314(6) & 0.153(7) &	0.113(7) & 0.072(6) & 3.986(43) & 8.162(64) & 6.128(103) & 3.1 \\
V15	& 18.883(3) & 0.300(5) & 0.139(5) & 0.094(4) & 0.066(4) & 3.907(46) & 8.137(72) & 6.222(98) & 3.9 \\
V16 & 18.858(7) & 0.369(10) & 0.197(9) & 0.150(10) & 0.099(10) & 3.885(70) & 8.186(99) & 5.899(141) & 4.3 \\
V17 & 18.934(4) & 0.384(6) & 0.176(6) & 0.132(6) & 0.074(5) & 3.750(45) & 7.769(63) & 5.654(100) & 3.4 \\
V18 &18.955(4)& 0.277(6)& 0.131(6)& 0.063(6)&  0.061(6)&  4.013(62)&  8.205(114)&  6.101(127)& 4.5\\
V20 & 18.961(3) & 0.305(4) & 0.151(4) &	0.109(4) & 0.058(4) & 3.978(39) & 8.399(55) & 6.497(91) & 1.8 \\
V21 & 18.910(3) & 0.231(5) & 0.104(5) & 0.081(5) & 0.041(5) & 4.177(63) & 8.721(87) & 7.025(145) & 3.7 \\ 
V22 & 18.935(5) & 0.375(7) & 0.163(7) & 0.128(8) & 0.087(8) & 3.909(61) & 8.016(77) & 6.000(106) & 2.9 \\
V23	& 18.797(6) & 0.286(9) & 0.146(8) & 0.090(6) & 0.053(7) & 3.971(87) & 8.208(143) & 5.777(193) & 3.6 \\
V24 & 18.356(2) & 0.133(3) & 0.067(3) & 0.038(3) & 0.017(3) & 4.058(68) & 8.522(110) & 6.766(203) & 4.1 \\
V25 & 18.844(3) & 0.313(4) & 0.139(4) & 0.116(5) & 0.077(4) & 3.961(43) & 8.157(56) & 6.304(82) & 1.6 \\
V27 & 18.257(28)& 0.219(40)& 0.102(13)& 0.052(30)& 0.036(26)&  4.385(590)& 8.781(805)&  6.254(1.000)& 8.7\\
V29 & 18.975(4) & 0.321(6) & 0.171(6) & 0.122(6) & 0.080(6) & 3.966(53) & 8.082(76) & 6.222(106) & 2.3 \\
V30 & 18.634(4)&  0.239(5)& 0.107(5)& 0.087(5)& 0.047(5)&  3.761(64)& 8.140(88)& 5.908(132)& 3.0\\ 
V31 & 18.901(6) & 0.374(9) & 0.124(9) & 0.065(9) & 0.053(9) & 3.962(87) & 7.694(158) & 6.085(197) & 9.2 \\
V32	& 18.756(4) & 0.248(5) & 0.121(5) & 0.088(5) & 0.055(3) & 3.814(60) & 8.102(86) & 5.883(121) & 2.2 \\
V33 & 19.032(4) & 0.331(6) & 0.151(6) & 0.122(6) & 0.088(5) & 3.781(55) & 8.142(77) & 6.022(105) & 4.2 \\
V35 & 18.983(3) & 0.261(4) & 0.121(4) & 0.085(4) & 0.046(4) & 4.096(47) & 8.457(68) & 6.601(112) & 6.0 \\
V38 & 18.837(3) & 0.284(5) & 0.142(5) & 0.099(5) & 0.056(5) & 3.843(51) & 8.179(74) & 6.435(114) & 3.0 \\
V41 & 18.635(4) & 0.290(6) & 0.124(6) & 0.093(6) & 0.053(5) & 3.752(58) & 7.640(82) & 5.719(132) & 7.6 \\
V43 & 18.944(4) & 0.216(5) & 0.100(5) & 0.065(5) & 0.033(5) & 3.765(67) & 8.320(103) & 6.761(173) & 6.3 \\
V44 & 18.910(4) & 0.291(5) & 0.117(6) & 0.077(6) & 0.037(5) & 3.965(63) & 8.166(97) & 6.195(178) & 10.9 \\
V45 & 18.903(3) & 0.258(4) & 0.119(4) & 0.094(4) & 0.054(4) & 3.985(49) & 8.378(69) & 6.440(106) & 4.4 \\
V46 & 18.872(4) & 0.120(5) & 0.024(6) & 0.015(6) & 0.004(6) & 5.010(295) & 7.852(446) & 5.724(1710) &20.5\\
V47 & 18.949(4) & 0.273(5) & 0.084(6) & 0.044(6) & 0.007(6) & 3.953(77) & 8.710(140) & 7.822(828) & 29.3 \\
V48 & 18.967(3) & 0.224(5) & 0.136(5) & 0.075(5) & 0.040(5) & 3.918(55) & 8.471(89) & 6.839(143) & 2.4 \\
V49 & 18.984(4) & 0.287(6) & 0.140(7) & 0.093(6) & 0.054(6) & 3.877(59) & 8.218(91) & 6.396(139) & 4.8 \\
V50 &18.919(4)& 0.297(5)&  0.154(5)&  0.105(5)&  0.071(5)&  3.807(49)& 8.217(72)&  6.262(101)& 4.3 \\
V51 & 18.977(3) & 0.156(4) & 0.055(4) & 0.036(4) & 0.016(4) & 4.283(88) & 8.799(130) & 7.095(279) & 46.8 \\
V52 &18.938(3)& 0.221(5)&  0.089(5)&  0.049(5)&  0.029(5)&  4.036(64)&  8.550(108)&  6.901(108)&1.8 \\
V56 & 19.086(7) & 0.391(10) & 0.157(9) & 0.147(8) & 0.079(9) & 3.650(79) & 8.195(101) & 5.826(153) & 6.8 \\
V58 & 18.851(3) & 0.106(5) & 0.029(5) & 0.006 (5) & 0.008(5) & 3.762(180) & 8.572(800) & 7.560(607) & 32.1 \\
V59 & 18.975(8) & 0.392(13) & 0.180(12) & 0.110(11) & 0.066(11) & 3.934(85) & 7.946(141) & 5.616(215) & 3.2 \\
V60 & 18.913(4) & 0.220(6) & 0.096(6) & 0.045(7) & 0.028(6) & 4.191(85) & 8.439(141) & 6.766(225) & 1.8 \\
V61 &18.817(5)& 0.369(7)& 0.168(7)&0.086(8)& 0.042(9)&4.163(57)&  8.355(100)&  5.796(172)& 4.3\\
V63& 18.801(7)& 0.354(10)& 0.147(10)& 0.091(10)&  0.053(10)&4.085(89)& 8.325(139)& 6.089(222)&4.0\\
V65 & 18.776(4) & 0.221(6) & 0.074(6) & 0.056(6) & 0.027(6) & 3.940(95) & 8.112(134) & 6.501(235) & 6.4 \\
V67&18.508(4)&0.147(6)& 0.082(6)& 0.033(6)  &0.009(5)& 4.251(103)& 9.072(205)&  9.271(634)& 42.4 \\
V68 & 19.015(7) & 0.296(10) & 0.175(10) & 0.128(10) & 0.049(10) & 4.072(91) & 8.401(130) & 6.709(245) & 8.1 \\
V69&19.422(14)&0.367(20)& 0.370(19)&  0.049(20)& 0.048(20)&  5.061(120)& 7.556(422)& 5.110(451)& 22.6\\
V73&18.129(4)& 0.173(5)&  0.109(5)& 0.052(5)&0.018(5)&3.915(79)& 8.252(144) & 5.891(317)& 7.8 \\
V76 & 19.118(8) & 0.446(12) & 0.221(12) & 0.118(12) & 0.027(12) & 4.297(77) & 9.032(138) & 6.932(483) & 11.1 \\
\hline
\multicolumn{10}{c}{RRc} \\
\hline
V53	& 18.837(3) & 0.103(4) & 0.014(4) & 0.020(4) & 0.004(4) & 3.013(300) & 5.230(226) & 3.169(884) & \\
V57 & 18.956(4) & 0.305(5) & 0.035(5) &	0.033(5) & 0.006(5) & 4.894(162) & 4.041(166) & 1.904(813) & \\
V62 & 18.873(4) & 0.211(5) & 0.028(6) & 0.029(5) & 0.004(5) & 4.646(174) & 2.827(187) & 1.485(1152)& \\
V64 & 18.877(4) & 0.179(5) & 0.046(5) &	0.021(5) & 0.011(5) & 3.438(124) & 1.770(253) & 5.366(474) & \\
V71 & 18.703(4) & 0.205(6) & 0.045(5) & 0.024(5) & 0.013(5) & 4.443(145) & 1.300(266) & 5.283(449) & \\
\hline
\multicolumn{10}{l}{The numbers in parenthesis indicate the uncertainty on the last decimals.} \\
\end{tabular}
\label{tab:coeficientes}
\end{table*}

The calibrations that correlate the Fourier and the physical parameters for RRab and RRc stars and their zero points, have been described in detail by \citet{Arellano2022} and in many of the references cited there, hence we will not repeat them here. The calibrations and their zero points can also be found summarized in Table 4 of \citet{Yepez2022}. We should recall however, that the [Fe/H] calibration for RRab stars \citep{Jurcsik1996}, is valid only for stars with a defined consistency parameter $D_m < 3.0$. In order to include a more solid sample we have slightly relaxed this condition and we have taken star with $D_m < 5.0$.

In Table \ref{tab:coeficientes} we provide the Fourier coefficients of all stars measured by our photometry, however, not all the stars were considered for the physical parameters calculation due to several reasons; first, if $D_m > 5.0$, or if the star position on the CMD is noticeable peculiar mostly due to unresolved blends, or if the membership analysis and the resulting distance suggest that the star is a field star. The peculiarities of individual stars are presented in Appendix A.

In Table \ref{tab:param} we list the resulting physical parameters only for stars with consistent light curves and likely cluster members.

\subsection{The mean metallicity and distance}

Columns 2 and 3 of Table \ref{tab:param} contain the iron abundance value in the scale of \citet{Zinn1984} [Fe/H]$_{\rm ZW}$ and its transformation to the {\it Ultraviolet and Visual Echelle Spectrograph}
(UVES) scale of \citet{Carretta2009} [Fe/H]$_{\rm UVES}$ respectively, correlated  via the equation,

\begin{equation}
{\rm [Fe/H]}_{\rm UVES} = -0.413 + 0.130{\rm [Fe/H]}_{\rm ZW} - 0.356{\rm [Fe/H]}_{\rm ZW}^2 .
\end{equation}

For completeness, column 4 of Table 3 lists the value given by the calibration of \citet{Nemec2013} scale, [Fe/H]$_{\rm Nem}$, also in the spectroscopic metallicity scale. 

The estimated weighted mean values are also indicated in the bottom of the table. For [Fe/H]$_{\rm UVES}$ we found  $-1.41\pm0.13$ and $-1.75\pm0.27$ from RRab and RRc stars respectively.

The mean cluster distance is estimated from the individual values of the absolute magnitude $M_V$ and the corresponding distances for a reddening $E(B-V)=0.08$. An average distance of $41.1\pm 1.5$ kpc is found from RRab stars and a very similar value for RRc stars.

\subsubsection{Cluster distance from {\rm P-L}(I) relationship}

An independent instrument to calculate the distance of globular cluster with RR Lyrae stars is the P-L relationship in the $I$ band of \citet{Catelan2004};
$M_I=0.471-1.132 \log P + 0.205 \log Z$, where $P$ is the fundamental period, $\log Z = {\rm [M/H]} - 1.765$, ${\rm [M/H] = [Fe/H]} - \log(0.638f + 0.205)$ and $\log f = {\rm [\alpha/Fe]}$ \citep{Salaris1993}.

For the calculation of the distance via P-L($I$) relation we have employed the intensity weighted mean magnitude $\langle I\rangle$ (see Table \ref{tab:periods}), $E(B-V)=0.08$ and $E(V-I)=1.259 E(B-V)$. We applied this P-L equation to 44 RRab stars given their fundamental period $P_0$. Four RRc stars were also included after fundementalizing their period assuming the ratio $P1/P0=0.749$. The resulting mean distance for these 48 RR Lyrae is $40.2\pm1.2$ kpc, in quite a good agreement with the
results from the Fourier decomposition (Table \ref{tab:param}).


\begin{table*}
\caption{Physical parameters for the RR Lyrae stars calculated using Fourier decomposition parameters and well-defined empirical calibrations. }
\centering
\begin{tabular}{cccccccccc}
\hline
Variable & [Fe/H]$_{\rm ZW}$ & [Fe/H]$_{\rm UVES}$ & [Fe/H]$_{\rm Nem}$& $M_V$ & $\log T_{\rm eff}$ & $\log(L/L_{\odot})$ & $D$ (kpc) & $M/M_{\odot}$ & $R/R_{\odot}$ \\
\hline
\multicolumn{10}{c}{RRab} \\
\hline
V1 & -1.49(5) & -1.40(6) & -1.28 & 0.64(1) & 3.831 & 1.646(3) & 40.13(13) &	0.61(1) & 4.86(2) \\
V2$^a$ &-- & -- & -- & 0.55(1) & 3.810 & 1.681(3) & 42.09(16) & 0.67(1)	& 5.59(2) \\
V3 & -1.39(7) & -1.29(7) & -1.12 & 0.59(1) & 3.793 & 1.662(3) &	41.72(14) &	0.83(1) & 5.90(2) \\
V5 & -1.55(10) & -1.48(11) & -1.46 & 0.64(1) & 3.820 &	1.643(4) & 38.87(19) & 0.61(1) & 5.09(2) \\
V6$^a$& --& --& --&  0.62(1)&  3.812 & 1.653(4)& 42.47(21)& 0.77(1)& 5.35(3) \\
V10$^a$	& -- & -- & -- & 0.52(1) & 3.819 & 1.691(4) & 43.10(20) & 0.70(1) & 5.42(3) \\
V11$^a$ & -- & -- & -- & 0.60(1) & 3.797 & 1.660(3) & 41.91(14) & 0.76(1) & 5.79(2) \\
V12 &-1.94(10) &-2.01(14) &-2.42&0.61(1)&  3.803 & 1.657(3)&40.65(16)&0.70(1)&5.61(2)\\
V13 & -1.65(8) & -1.60(9) & -1.67 &	0.61(1) & 3.809 & 1.655(4) & 41.56(19) & 0.69(1) & 5.43(2) \\
V14 & -1.53(6) & -1.44(7) & -1.39 & 0.61(1) & 3.798 & 1.656(4) & 41.66(17) & 0.76(1) & 5.72(2) \\
V15 & -1.66(7) & -1.61(8) & -1.66 &	0.57(1) & 3.795 & 1.671(3) & 40.98(12) & 0.77(1) & 5.91(2) \\
V16 & -1.42(9) & -1.32(10) & -1.18 & 0.61(1) & 3.795 & 1.655(6) & 39.77(26) & 0.84(1) & 5.80(4) \\
V17 & -1.71(6) & -1.68(8) & -1.81 & 0.62(1) & 3.821 & 1.652(3) & 41.03(16) & 0.66(1) & 5.12(2) \\
V18&-1.65(11)& -1.60(13)& -1.63 &0.55(1)&
3.796 & 1.679(3)&42.76(17)&0.75(1)&5.95(2)\\
V20 & -1.37(5) & -1.26(5) & -1.06 & 0.59(1) & 3.811 & 1.662(2) & 42.06(11) & 0.64(1) & 5.42(1) \\
V21 & -1.20(8) & -1.08(7) & -0.71 & 0.61(1) & 3.787 & 1.657(3) & 40.82(13) & 0.76(1) & 6.02(2) \\
V22 & -1.54(7) & -1.46(8) & -1.42 & 0.60(1) & 3.815 & 1.659(4) & 41.38(20) & 0.69(1) & 5.31(3) \\
V23 & -1.66(13) & -1.61(16) & -1.65 & 0.56(1) & 3.805 & 1.676(5) & 39.61(21) & 0.66(1) & 5.67(3) \\
V25 & -1.47(5) & -1.37(6) & -1.27 & 0.64(1) & 3.849 & 1.644(2) & 39.01(11) & 0.42(1) & 4.46(1) \\
V31$^a$ & -- & -- & -- & 0.50(1) & 3.749 & 1.701(5) & 42.75(25) & 1.54(2) & 7.57(4) \\
V32 & -1.59(8) & -1.52(9) & -1.53 & 0.66(1) & 3.822 & 1.634(3) & 37.05(12) & 0.54(1) & 4.99(2) \\
V33 & -1.53(7) & -1.45(8) & -1.41 & 0.61(1) & 3.785 & 1.658(3) & 43.26(17) & 0.91(1) & 6.10(2) \\
V35$^a$ & -- & -- & -- & 0.60(1) & 3.788 & 1.660(2) & 42.39(11) & 0.79(1) &	6.02(2) \\
V38 & -1.77(7) & -1.73(9) & -1.83 & 0.55(1) & 3.818 & 1.681(3) & 40.56(13) & 0.55(1) & 5.36(2) \\
V43$^a$ & -- & -- & -- & 0.64(1) & 3.803 & 1.646(3) & 40.95(13) & 0.64(1) & 5.54(2) \\
V45	& -1.41(7) & -1.31(7) & -1.14 & 0.63(1) & 3.788 & 1.649(2) & 40.33(11) & 0.80(1) & 5.96(2) \\
V46$^a$&-- &-- &--&0.61(1)&3.806 & 1.656(3)&40.06(14) &0.54(1)&5.51(2)\\
V47$^a$ & -- & -- & -- & 0.59(1) & 3.805 & 1.663(3) & 41.87(15) & 0.71(1) & 5.60(2) \\
V48	& -1.43(8) & -1.33(9) & -1.14 & 0.61(1) & 3.790 & 1.655(3) & 41.82(14) & 0.74(1) & 5.93(2) \\
V49 & -1.56(9) & -1.48(10) & -1.44 & 0.60(1) & 3.799 & 1.662(3) & 42.48(17) & 0.73(1) & 5.75(2) \\
V51$^a$ &-- & -- & -- & 0.62(1) & 3.774 & 1.653(2) & 41.91(11) &	0.82(1) & 6.36(2) \\
V52 &  -1.40(19)& -1.29(10)&-1.04&0.58(1)&  3.783 & 1.668(3)&41.86(14)&0.82(1)& 6.23(2)\\
V56 &  -1.45(10)& -1.35(10)& -1.23&0.57(1)&3.775&  1.673(5)&45.07(28)&1.09(2)&6.49(4)\\
V58 & -1.84(75) &-1.85(98)& -1.63&0.53(1)&3.782 & 1.687(3)&41.12(13) &0.67(1)& 6.38(2)\\
V59 & -1.43(13) & -1.33(14) & -1.15 & 0.64(2) & 3.831 & 1.643(7) & 41.39(33) & 0.63(1) & 4.85(4) \\
V60 & -1.52(13) & -1.43(14) & -1.28 & 0.57(1) & 3.774 & 1.670(4) & 41.52(17) & 0.90(1) & 6.50(3) \\
V61 & -1.46(10)& -1.36(10)& -1.22&0.49(1)& 3.825&  1.706(4)& 41.37(20)&0.60(1)&5.36(3)\\
V63 &-1.33(13)& -1.21(13)& -1.00&0.57(1)& 3.833&  1.674(6)& 39.57(26)&0.55(1)&4.97(3)\\
V68$^a$ & -- & -- & -- & 0.65(1) & 3.783 & 1.640(6) & 42.01(27) & 0.89(1) & 6.03(4) \\
\hline
Weighted Mean & -1.53(1) & -1.43(1) & -1.41 & 0.597(1) & 3.802 & 1.660(1) & 41.17(22) & 0.67(1) & 5.61(1) \\
$\sigma$ & $\pm 0.15$ & $\pm 0.20$ & $\pm 0.33$ & $\pm 0.040 $ & $\pm 0.020$ & $\pm 0.016$ & $\pm 1.39$ & $\pm 0.18$ & $\pm 0.56$ \\
\hline
\multicolumn{10}{c}{RRc} \\
\hline
V53 & -1.59(34) & -1.52(38) & -- & 0.73(2) & 3.882 & 1.609(9) & 37.33(38) & 0.50(1) & 3.68(4) \\
V57 & -1.52(34) & -1.44(37) & -1.43(27) & 0.55(2) & 3.853 & 1.681(9) & 42.87(47) & 0.55(2) & 4.57(5) \\
V62 & -1.99(36) & -2.08(51) & -2.01(24) & 0.57(2) & 3.827 & 1.671(9) & 40.78(45) & 0.74(2) & 5.09(6) \\
V64 & -1.99(43) & -2.09(61) & -1.83(25) & 0.64(2) & 3.879 & 1.646(9) & 39.70(42) & 0.45(1) & 3.90(4) \\
V71 & -2.09(46) & -2.24(67) & -2.03(25) & 0.55(2) & 3.833	& 1.679(9) & 38.05(41) & 0.78(2) & 5.00(5) \\
\hline
Weighted Mean & -1.79(16) & -1.72(21) & -1.84(10) & 0.61(1) & 3.853 & 1.656(4) & 39.49(19) & 0.55(1) & 4.27(2) \\
$\sigma$ & $\pm 0.27$ & $\pm 0.27$ & $\pm 0.28$ & $\pm 0.07 $ & $\pm 0.022$ & $\pm 0.027$ & $\pm 2.22$ & $\pm 0.13$ & $\pm 0.57$ \\
\hline
\multicolumn{10}{l}{$a$: These variables are not included in the calculation of [Fe/H] since their $D_m$ parameter is larger than 5.0} \\
\multicolumn{10}{l}{The numbers in parentheses indicate the uncertainty on the last decimal place. } \\
\end{tabular}
\label{tab:param}
\end{table*}

\section{The Colour Magnitude Diagram of NGC 7006}

\begin{figure*}
\centering
\includegraphics[width=0.99\textwidth, trim = 0cm 0cm 0cm 0cm]{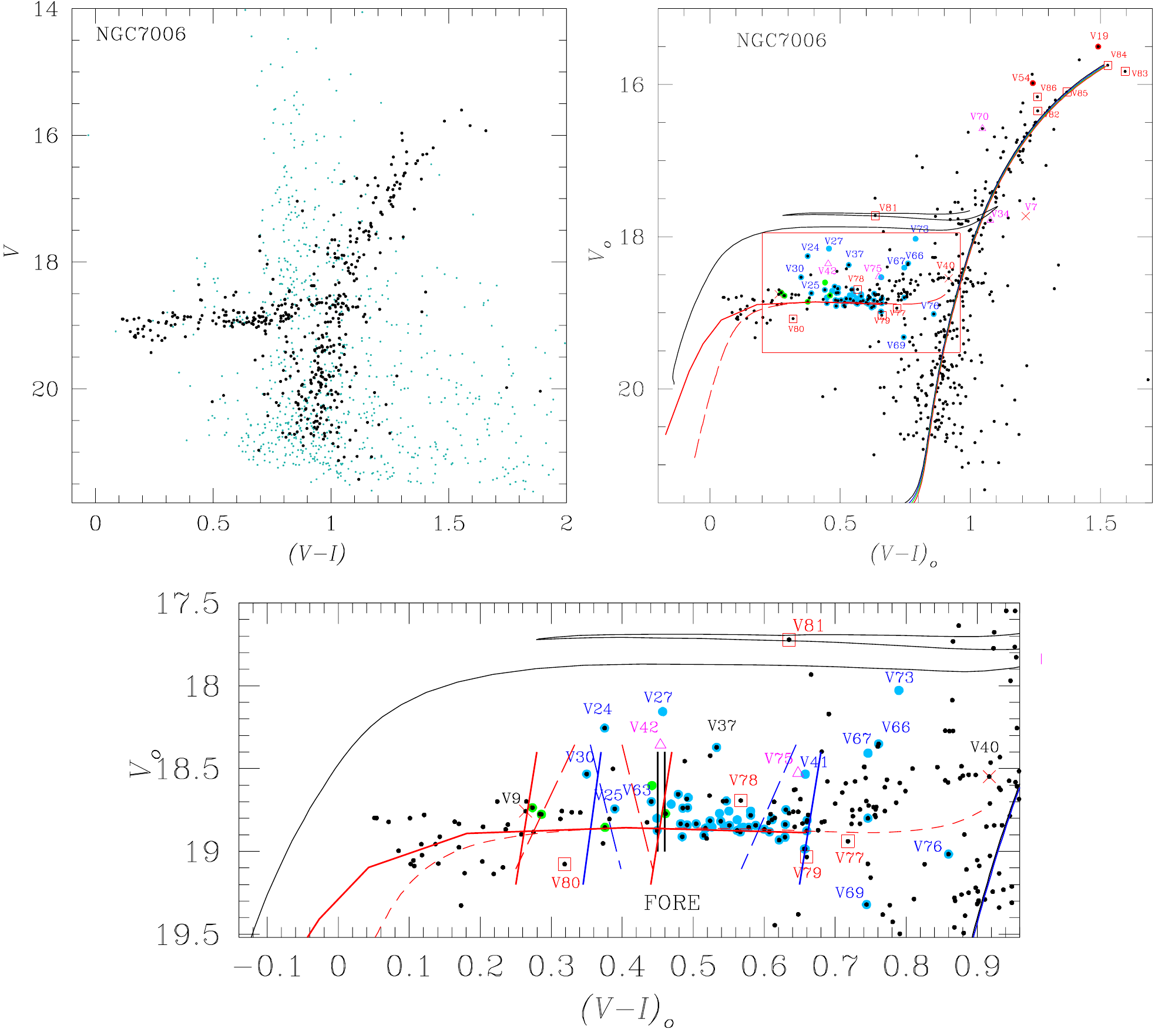}
\caption{CMD of NGC 7006 and the variable stars distribution. The top left panel shows the all stars measured by our photometry. Light green and black dots represent the field stars and the likely cluster members respectively ($\$$ \ref{MEMBERSHIP}). The top right panels is the unreddened CMD with the variable stars identified. The dashed ZAHB and isochrones for 12.0 to 13.5 Gyrs and [Fe/H]=-1.58, $Y = 0.25$ and [$\alpha$/Fe]=+0.4 are from the models of \citet{Vandenberg2014}. The continuum ZAHB is from the models
built from the Eggleton code \citep{Pols1997, Pols1998, KPS1997}, and calculated by \citet{Yepez2022}. The black continuous line is an evolutionary track of a star starting in the blue tail of the HB, with a core and total masses of 0.49 and 0.55 $M_{\odot}$ respectively. This track represents well the BL Her star V81. All theoretical loci were place at a distance of 44 kpc that was found to provide the best fits. The red square is magnified in the bottom panel where the variable stars are coded as: RRab blue, RRc green, SR red, magenta triangles are for stars of confirmed variability in this work, crosses for variables not confirmed by our photometry. Red squares mark the newly found variables. Tilted continuous blue and red lines represent our empirical estimates of the fundamental mode and the first overtone instability strip edges respectively. For comparison the theoretical lines from \citet{Bono1994} (their Figure 4) are shown with broken lines (see the text for details). The vertical black lines are the empirical First Overtone Red Edge (FORE) from \citet{Arellano2015,Arellano2016}.}
\label{DCM}
\end{figure*}

Fig. \ref{DCM} displays the CMD of NGC 7006 built based on the \emph{VI} photometry of the work. The top left panel display the distribution of all stars measured in the field of NGC 7006, and distinguishes the field and member stars resulting from the analysis of $\S$ \ref{MEMBERSHIP}. The top right panel displays the variable stars distribution on the unreddened plane $(V-I)_0 - V_0$. All theoretical loci have been placed at a distance of 44.0 kpc, since we found this to produce the best representation of the stellar distribution. We should note the small difference with the distance obtained from the Fourier RR Lyrae light curve decomposition (Table \ref{tab:param}) and from the P-L(I) relation, which is 40.2$\pm$1.2 kpc. The difference is the consequence of the RR Lyrae being  intrinsically brighter than ZAHB due to evolutionary reasons. The bottom panel of Fig. \ref{DCM} is an expansion of the HB region that helps studying in detail the distribution of the RR Lyrae stars. Several aspects should be highlighted from this plot: first, we note that all variables (coloured symbols as coded in the figure caption) matching a cluster member star (black points) carry a black dot in the center, the absence of which means that the star was found to be likely a field star by our approach of $\S$ \ref{MEMBERSHIP}, (see for example the cases of V27, V67 and V73 among others). Second, a few RRab and RRc stars are at an odd position, either too luminous for  a cluster member at the HB, or in the "wrong" side of the first overtone red edge (FORE). This could be due to the fact that the star is not a cluster member (e.g. V27 and V73) or that in our images the star is blended with a brighter star (e.g. V30, V69). In Appendix A we include comments on peculiar stars.  It is worth mentioning that all the non-peculiar RRab and RRc concentrate near the ZAHB and the two modes
are well separated by the FORE, i.e., no RRab stars are found in the multi-mode region just to the blue of the FORE. This circumstance, that is a common feature in OoII clusters, has been identified only in some OoI clusters \citep[e.g.][]{Yepez2022}, and this seems to be the case of NGC 7006. The solid tilted red and blue lines are  our empirical
estimates of the fundamental mode and the first overtone instability strip edges respectively. For comparison we have included the theoretical instability strip bounds calculated by \citet{Bono1994} (their Figure 4) converted from the log $T_{\rm eff}$ - log $L$ plane to the $(V-I)_0 - V_0$ observational plane. This was achieved by adopting the cluster distance 41.2 kpc and the   $T_{\rm eff}$ vs $(V-I)_0$ described by \citet{Arellano2010}. 

\section{On the progenitors of RR Lyrae and BL Her stars}

We would like to understand which stars evolved to become the RR Lyrae stars in NGC 7006.
Similarly we would like to explain the progenitor of stars like the BL Her type V81 in NGC 7006. For both purposes we have employed the Eggleton code \citep{Pols1997, Pols1998, KPS1997} and a modified mass loss Reimers law  \citep{SC2005} to produce ZAHBs models for different He core and envelope masses.

\begin{figure}
\centering
\includegraphics[width=1\columnwidth]{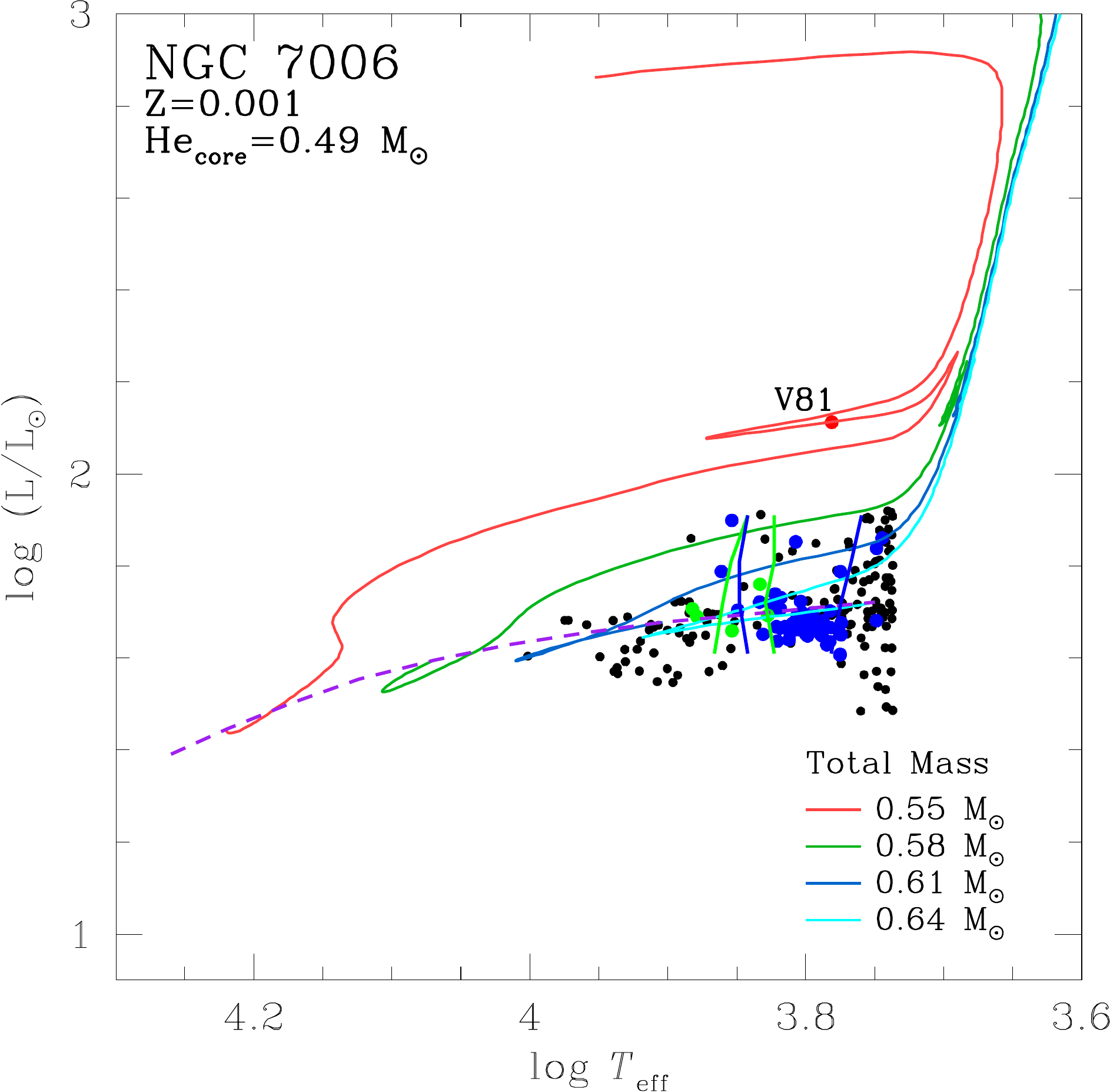}
\caption{HR diagram of the HB of NGC 7006. The evolutionary tracks were calculated for a core mass
of 0.49 $M_{\odot}$ and the total masses in the legend. Only configurations with thin a envelope (0.06 ) would have 
sufficient luminosity to represent the BL Her star V81. Massive envelopes also do not produce blue loops. Blue and green nearly vertical lines are the theoretical instability strip borders for the fundamental and first overtone respectively, taken from \citet{Bono1994}.  }
\label{HR_7006}
\end{figure}

Theoretical calculations of post-HB evolution have been carried out by several authors using different 
codes \citep[e.g.][]{Osborn2019,Bono2020,Yepez2022}. 
All these studies agree that 
globular cluster cepheids or BL Her stars, have their progenitors from the blue tail 
of the ZAHB as their He-core get exhausted, and may have total masses in the range 0.52-0.58 
$M_{\odot}$ for a canonical helium abundance of Y=0.25. \citet{Osborn2019}, studying
the secular period changes, found that in a sample of 18 BL Her stars, all have positive 
period changes, i.e. they are all evolving redwards, and that depending upon their period 
they may  be in the first redward crossing of the instability strip or the second redward crossing (after the He shell flashes in the AGB that produce the blue loops). No stars with blueward evolution 
(negative period changes) were found. Comparison with theory allow these authors to conclude that 
BL Her stars have helium abundances close to the canonical Y=0.25 and that values as large as 0.33 can be excluded. In the characterization of the evolution of BL Her and W Vir stars
made by \citet{Bono2020}, we learn that BL Her are post-early asymptotic giant branch (PEAGB) stars with masses 0.495 <$M_{\odot}$ < 0.55 
evolving along their first crossing towards the AGB, while W Vir stars are a mix of PEAGB and post-AGB stars (hydrogen shell burning). \citet{Yepez2022} have pointed out that models with $M > 0.54 M_{\odot}$ can represent BL Her stars
only if the mass of the envelope is small.

The continuous ZAHB in Fig. \ref{DCM} was obtained for models with a core mass of 0.49 $M_{\odot}$ and total mass in the range 0.55-0.64 $M_{\odot}$. Notice that larger core masses would naturally produce more luminous ZAHB's. The dashed ZAHB is from the models of \citet{Vandenberg2014} for an initial core mass of 0.49 $M_{\odot}$. Our models indicate that a main sequence star with a mass between 0.82 and 0.85 $M_{\odot}$ would reach the RGB in 12 to 13.5 Gyrs. When progenitors within this mass range reach the ZAHB they have lost between 25-35\% of their mass due to He-flash events at the RGB.
The continuous black line in Fig. 5, is an evolutionary track starting
at the blue end of the HB and rising to the position of the BL Her
star V81. This model has a core mass of 0.49 $M_{\odot}$ and a total mass of
0.55 $M_{\odot}$ , i.e. a very thin envelope of 0.06 $M_{\odot}$. 

Fig. \ref{HR_7006} shows the HR diagram of NGC 7006 and our 
models for a core mass of 0.49 $M_{\odot}$ and envelope masses in the range 0.06-0.15 $M_{\odot}$.
Models with massive envelopes will remain at luminosities much lower than that of the V81
and will not produce blue loops. 
We have also found for the cases of Pop II Cepheids in M14 \citep{Yepez2022} and M10 \citep{aaf20},  that these thin envelopes are necessary to explain BL Her and W Virginis with low mass predecessors from the blue tail of the HB.

\section{Summary of results}

Accurate \emph{VI} magnitudes were extracted for 1373 star in the field of NGC 7006. Using the $Gaia$-DR3 accurate sky positions and proper motions we have separated 470 stars that are likely cluster members. This enabled us to build a clean CMD and discuss its structure and further improve the membership status of the population of variable stars. In this process we have explored the light curves of all our photometric sources and discovered 10 new variables; 2 RRab, 2 RRc, 1 CWB and 5 SR.
The coordinates and classifications of a few variables were corrected, e.g. the cases of the two CW stars V70 and V81. A few variables in the FoV of the clusters on close scrutiny were found to be field stars.
The cluster is confirmed as of the Oo I type.

The $(V-I)$ colour curves of RRab stars is known to be constant. We have taken advantage of that to estimate the reddening of the cluster as 0.08$\pm$0.05. 
The light curves of the confirmed RR Lyrae members were Fourier decomposed and their physical parameter were calculated. We provide the Fourier parameters as well as the individual physical parameters, effective temperatures, masses and radii besides the metallicity and distance. The RRab stars lead to the mean cluster metallicity and distance of [Fe/H]$_{\rm ZW}= -1.53\pm0.15$ and $41.2\pm1.4$ kpc. 

The CMD diagram displays the distribution of variable members in good concordance with theoretical loci, i.e. ZAHB and isochrones placed a the resulting distance. We notice the presence of RRab stars in the multi mode-region of the HB, which is
a common feature in some Oo I clusters.

The employment of evolutionary models that use the Eggleton code, allowed us to estimate the mass of the HB stars predecessors at the main sequence to be in the range 0.82 and 0.85 $M_{\odot}$ and to infer that by the time these stars reach the HB, 12-13.5 Gyrs later they may have lost 25-35\% of the mass. Likewise, the models enabled us to demonstrate that CW stars (BL Her or W Virginis stars) can be explained by stars evolved from the blue tail of the HB provided that the envelope is thin ($\sim$ 0.06 $M_{\odot}$), or in other words with a large core to total mass ratio.
 

\section*{Acknowledgements}

AAF acknowledges the financial support of the program m PAPIIT, UNAM, through grant IG100620. FCRG is grateful to CONACyT (Mexico) for a M. Sc. scholarship. We thank the staff of the
IAO, Hanle and CREST, Hosakote, for making these observations
possible. The facilities at IAO and CREST are operated by the Indian
Institute of Astrophysics, Bangalore. We have made extensive use
of the SIMBAD and ADS services, for which we are thankful.


\section*{DATA AVAILABILITY}
The data underlying this article shall be available in an electronic
form in the Centre de Donnés astronomiques de Strasbourg data
base (CDS), and can also be shared on request to the corresponding
author.



\bibliographystyle{mnras}
\bibliography{N7006} 

\appendix
\label{Appendix}

\section{Comments on individual stars}

Here we address a few stars whose light curve, classification, identification, membership status or position in the CMD display some peculiarity.

 V7 and V40. These two star, identified with a cross in Fig. \ref{DCM}, do not show variations in our data. V7 is in fact labeled as a non variable in the CVSGC.
 
 V24. This RRab star appears as a likely cluster member, in spite of which it is too luminous in the DCM. Furthermore the P-L($I$) relation gives a distance of 31.8 kpc. Therefore the star is likely a foreground star.

V26, V37, V39, V55. These stars we not measured by our photometry because they are to faint and/or blended in a crowded region. V55 is out of out FoV.
 
 V27, V41 and V73. These stars are found to be field stars. Their position in the CMD is in fact peculiar, being too luminous to pertain to the HB, hence we consider this star as non cluster members. ~
 
\begin{figure}
\centering
\includegraphics[width=1\columnwidth]{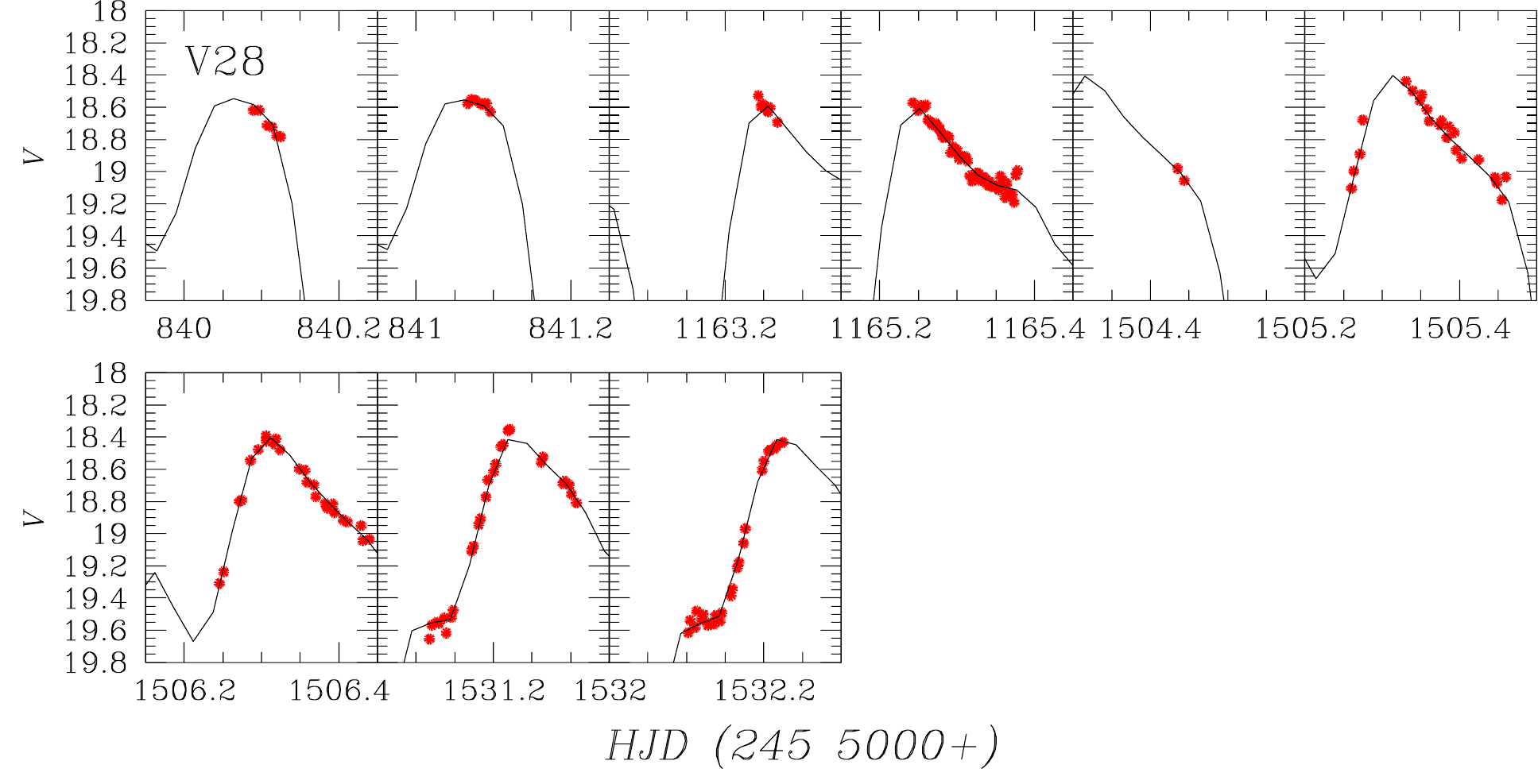}
\caption{V28 V data phased with a double mode model with periods 0.33213 and 0.49914 d.}
\label{V28_DM}
\end{figure}

 V28. This star has been classified as RRab in the CVSGC with a period of  0.4970 d. However our data are not phased at all by this period and rather suggest a period  of 0.3321 d. We have processed our data through a double mode model analysis and find two simultaneous active periods of 0.33213 and 0.49914 d which, if interpreted as the first overtone and fundamental periods have a ratio P1/P0= 0.665. This value is substantially smaller than the canonical 0.746 ratio \citep{Cox1983} for RRd stars. It seems clear that the star is double mode pulsator, but our data are rather insufficient for a more accurate evaluation of the involved modes, periods and their ratio. In Fig. \ref{V28_DM} the $V$ data of V28 are fitted by a double mode wave of the above periods. The representations is quite satisfactory but we recognize that our data are scarce for further analysis.
 We can only comment at present that, although the period ratio 0.665 is larger than in the recently discussed double mode stars with 
excited harmonics of non-radial modes, 0.61 \citep{Netzel2022}, the star is worth an analysis 
when more observations of good quality are available. 
 
V30. The star is listed in the CVSGC as non variable. We found the coordinates listed there were off by a few arc seconds and corrected the coordinates. At the correct position we find a clear RRab. The star is nevertheless too luminous in the DCM and we doubt its membership status. Hence, it was not used for the Fourier analysis.

V34. Related with the controversy on the identification and variability of this star, the reader can refer to the note in the CVSGC \citep{Clement2001} (2015 edition). Here we were guided by the coordinates given in this catalogue and the identification chart of \citet{Wehlau1992}, that in fact coincide. This star is obviously brighter than cluster RR Lyrae stars by more than one magnitude. In our data the star presents a sinusoidal mild
variation  with period of 0.9088 days. 
In the CMD it is found in the RGB. The membership analysis identifies it as a cluster member. In our opinion the star may be a binary star probably of the EW type.

V36, V66, V72. These three stars form a group and are badly blended in our images. We could only isolate a reasonable light curve for V66. In the CVSGC the quoted period for V66 is 0.6172 d, however we found it to be better phased with 0.3557 d, suggesting we area dealing with a RRc star. Nevertheless, the mutual light contamination of this group makes them all unsuitable for proper analysis. Thus we refrain from further comments on the nature of V66. This circumstance also explains the odd position of V66 in the DCM.

V42. This star is very close to the cluster center in a crowded region. It is near to a much brighter star. The present data are apparently the first ever published. Our light curve is scattered, nonetheless the period suggest the star being an RRab. The membership analysis finds the star to be a field star.

 V60, V67. These star were classified as RRc stars in the CVSGC. We found periods and light curve shapes typical of RRab stars hence we reclassified them. V67 appears to red and too luminous relative to the HB. It is also found not to be a cluster member and it distance via the P-L($I$) relation is much too short.

V68, V69. These stars had no Bailey type in the CVSGC. We found periods and light curve shapes typical of RRab stars hence we reclassified them. V69, in spite of having been selected as a likely
cluster member by the membership analysis, in the DCM appears much below the HB. Also, the
P-L($I$) relation renders a distance of 50.3 kpc, hence the star seems to be behind the cluster.

V70. This star carries a note in the CVSGC indicating that the star is probably not a RR Lyrae star \citep{PintoRosin1973}. Its position on the CMD is much brighter than the HB and towards the RGB. We find the star to be a cluster member. Our data are properly phased with a period of 11.7 d. Given its position on the CMD and its period the star could well be a W Virginis or CW star.

\begin{figure}
\centering
\includegraphics[width=1\columnwidth]{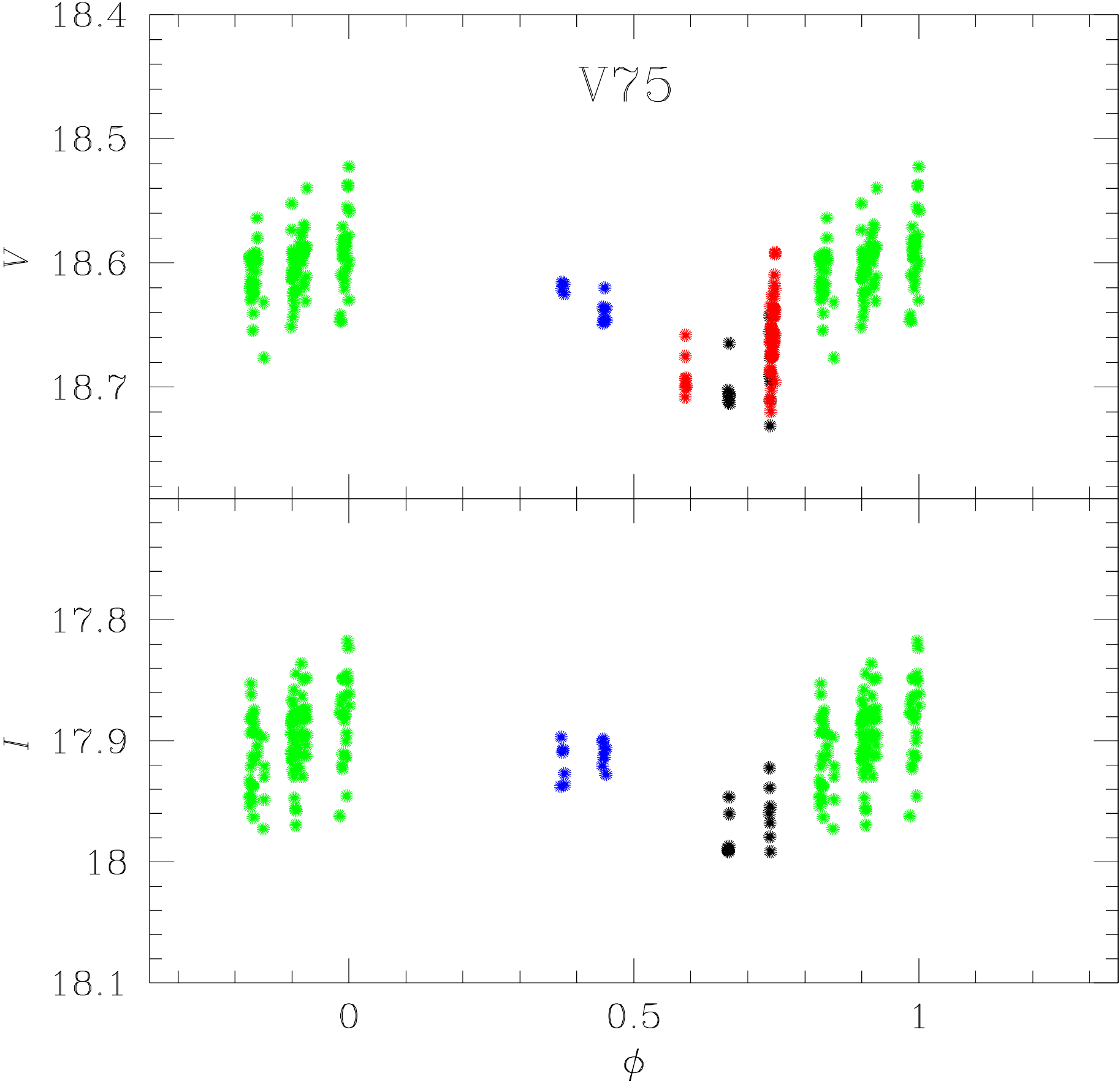}
\caption{V75 phased with P=13.5071 d, which along with its position in the DCM suggests the star to be a W Virginis (CWA) slightly behind the cluster.}
\label{V75}
\end{figure}

V75. No short term variations were detected for this star by \citet{Wehlau1999}. Our light curve is noisy but in the V vs HJD plane it displays clear variations in both $V$ and $I$ bands, with an amplitude of about 0.1 mag. A period search finds 13.5071 d. and the phased light curves can be seen in Fig. \ref{V75}. W agree with \citet{Rosino1967} that that it is not an RR Lyrae. Its position slightly above the the HB and its period suggest that it is a CWA or W Virginis star somewhat behind the cluster. 

V76.  This faint RRab may be subject to light and colour contamination from it brighter neighbour V31. The star appears too red and fainter than most RRab's on the HB which may indicate that the star does not belong to the cluster in spite to have been identified as a likely member by our membership analysis.

V1, V11, V12, V13, V14, V16, V27, V41, V42, V44, V45, V46, V47, V65, V67, V71, V73, V75.  All these stars are found as likely field stars by the membership analysis of $\S$ \ref{MEMBERSHIP}. However, while the method is very proficient in identifying the high probability cluster members as a whole, individually, the members and field stars populations can be mutually contaminated. Therefore, before labeling these variables as field stars we have taken into account other aspects of our analysis, e.g. to be considered a likely member, first, the star position of the CMD should not be obviously peculiar, as several howlers are obvious in the Fig. \ref{DCM}, some of which have been addressed in the above paragraphs. And second, the distance obtained from the absolute magnitude Fourier decomposition calibration and the one obtained via the P-L($I$) relation should be similar to the general mean distance within $\pm$3 kpc which is about 2 sigma. Applying these considerations we are left only with the following stars V27, V41, V42, V67, V71, V73 and V75 as most likely field stars.








\bsp	
\label{lastpage}
\end{document}